\documentclass[prc,showpacs,floatfix]{revtex4}
\usepackage{bm}
\usepackage{dcolumn}
\usepackage{graphicx}
\usepackage{amsmath}
\newcommand{\beq}{\begin{equation}}
\newcommand{\eeq}{\end{equation}}
\newcommand{\beqa}{\begin{eqnarray}}
\newcommand{\eeqa}{\end{eqnarray}}
\begin{document}

\title{Complete set of observables for photoproduction of two pseudoscalars on a nucleon}
\author{
H.~Arenh\"ovel$^{1}$\thanks{\emph{eMail address:} arenhoev@kph.uni-mainz.de} and
A.~Fix$^{2}$\thanks{\emph{eMail address:} fix@tpu.ru}}
\affiliation{\mbox{$^1$Institut f\"ur Kernphysik, Johannes Gutenberg-Universit\"at Mainz,
Mainz, Germany}\\
\mbox{$^2$Tomsk Polytechnic University, Tomsk, Russia}}

\date{\today}

\begin{abstract}
The problem of determining completely the spin amplitudes of
photoproduction of two pseudoscalar mesons on a nucleon from
observables is studied. The procedure of reconstruction of the
scattering matrix elements from a complete set of observables is
based on the expressions of all observables as quadratic
hermitean forms in the reaction matrix elements which are derived
explicitly. Their 
inversion allows one to find explicit solutions for the  reaction
matrix elements in terms of observables. Two methods for finding a
complete set of observables are presented. In particular, one set was
found, which does not contain a triple polarization observable. 
\end{abstract}


\maketitle

\section{Introduction}\label{introduction}

Experiments which are presently conducted at MAMI, ELSA, CEBAF, and
other research centers yield a large amount of new, very precise data
on meson 
photoproduction on nucleons. This has awakened renewed interest in
a comprehensive theoretical analysis of these reactions. The main
purpose of this study is to get unambiguous quantitative information
on the reaction amplitudes. An obvious method of solving this task
is a model independent analysis of a complete set of
measurements. Since the observables are nonlinear (quadratic)
functions of the amplitudes, the number of linearly independent forms
of observables generally exceeds the number of the amplitudes
sought. Thus a challenging task is to find a minimal set of
observables, i.e.\ a so-called complete experiment, which on the one
hand allows one to unambiguously determine the reaction amplitudes,
and on the other hand whose measurement is technically as
simple as possible.

Concerning reactions in which two pseudoscalar mesons are
produced, one faces at present quite an unusual situation insofar as a
large 
amount of precise data exists, in particular on polarization observables,
but only few theoretical studies are devoted to this problem. Among the
latter one is the work of Roberts and Oed~\cite{RoO05} where general
expressions for polarization observables in terms of helicity and
transversity amplitudes were obtained. Recently, the problem of a
truncated partial wave analysis of a complete experiment for such type
of reaction was considered in detail in~\cite{FiA13}.

As was noted in Refs.~\cite{RoO05} and \cite{FiA13}, in order to
determine all spin amplitudes for the photoproduction of two spin-zero
pseudoscalar mesons (up to an overall phase) one needs at least 15
observables. Such a minimal set of linearly independent observables is
called a ``complete set'',  which however may suffer from so-called
discrete ambiguities. The question of such a complete set
was already addressed in Ref.~\cite{RoO05}, where it was pointed out
that it will contain at least one triple polarization observable. 

The present paper is devoted to a mathematical solution 
of the problem of finding a complete set of observables for reactions
in which two pseudoscalar mesons are produced on a nucleon. In
particular, we have obtained expressions
which allow one to determine all photoproduction amplitudes if the
required minimal set of observables is known.
In the next two sections we review the general expressions of
Ref.~\cite{FiA12} for the reaction matrix and the various observables
which determine the most general differential cross section including
beam and target polarization and the target nucleon recoil
polarization. In Sect.~\ref{completesets} we present two methods,
allowing an explicit construction of a complete set of
observables. Here we also address a question,
concerning the elimination of triple polarization observables from a
complete set. Some formal ingredients and details
are collected in Appendices A to D. 

\section{The $T$-matrix}\label{tmatrix}
\begin{figure}[h]
\begin{center}
\includegraphics[scale=.8]{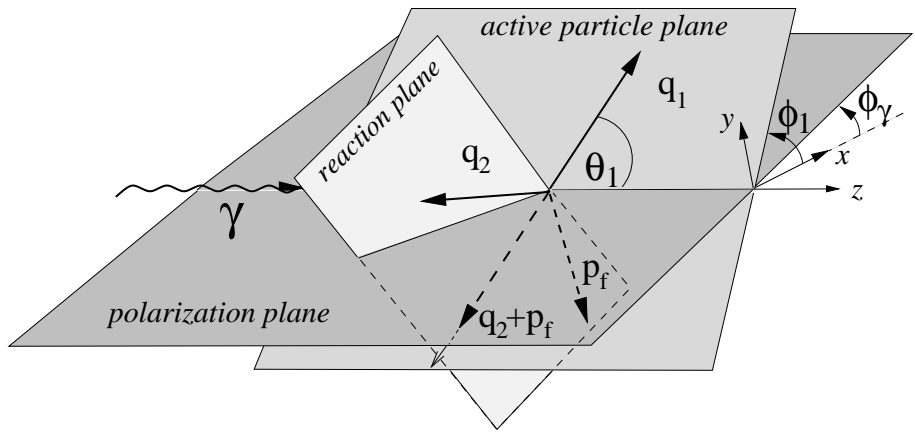} \caption{Definition of the coordinate frame
in the c.m.\ system. } \label{fig1}
\end{center}
\end{figure}

All observables are determined by the reaction or $T$-matrix. Its
specific form depends on the reference frame. Thus we will briefly
review the framework adopted in Ref.~\cite{FiA11} for the
photoproduction of two pseudoscalar mesons on a nucleon, namely 
$\eta$ and $\pi$. Cross section and recoil polarization are defined
with respect to the overall c.m.~system. With respect to this system
the four-momenta of incoming photon, outgoing two mesons, initial and
final nucleons are denoted by $(\omega_\gamma,\vec{k}\,)$,
$(\omega_1,\vec{q}_1\,)$, $(\omega_2,\vec{q}_2\,)$,
$(E_i,\vec{p}_i\,)$, and $(E_f,\vec{p}_f\,)$, respectively. The
definition of the reference frame is shown in Fig.~\ref{fig1}. The
$z$-axis is taken along the incoming photon momentum and $x$- and
$y$-axes are chosen arbitrarily to form a right-handed coordinate
system. In the case of linearly polarized photons the direction of linear
polarization defines another plane, the ``polarization plane'' with an
angle $\phi_\gamma$ with respect to the $x$-$z$-plane. 
Meson ``1'' with momentum 
$\vec q_1=(q_1,\Omega_1)$ is called the active particle. Its momentum
together with the photon momentum defines the ``active particle
plane'' which is inclined by an angle $\phi_1$ with respect to the
$x$-$z$-plane.  Furthermore, the momenta of final three
particles  define a plane which we call the ``reaction plane''.

We choose as independent variables for the description of this reaction the photon energy
$\omega=k$, the momentum of the outgoing active particle $\vec q_1$ and the spherical
angles $\Omega_{p}=(\theta_{p},\phi_{p})$ of the relative momentum $\vec p$ of the
outgoing meson ``2'' and nucleon as given by 
\beq 
\vec p=(M_p\vec q_2-m_2\vec
p_f)/(M_p+m_2)=(p,\Omega_{p})\,.\label{momentum-p} 
\eeq 
The momentum $\vec p$ is located in the reaction plane. 
Then the momenta of the second meson and the outgoing nucleon are
fixed. For example, the meson momentum reads 
\begin{equation}\label{qp}
\vec q_2=\vec p-\frac{m_2}{M_p+m_2}\,{\vec q_1}\,.
\end{equation}
In the following we will use for the active particle $\vec
q=(q,\Omega_q)$ instead of $\vec q_1$ for convenience.

In Ref.~\cite{FiA11} the following expression for the $T$-matrix had
been derived by expansion of the final state into partial waves
\begin{eqnarray}\label{small_t}
T_{m_f \mu m_i}(\Omega_{p},\Omega_q)&=& e^{i(\mu+m_i-m_f)\phi_q} t_{m_f \mu
m_i}(\theta_{p},\,\theta_q,\,\phi_{pq})\,,
\end{eqnarray}
allowing the separation of the $\phi_q$-dependence such that
the small $t$-matrix elements depend on $\theta_{p}$, $\theta_q$,
and the relative azimuthal angle $\phi_{pq}=\phi_p-\phi_q$ only. The
spin quantum numbers
$\mu$, $m_i$, and $m_f$ refer to the photon and initial and final
nucleon, respectively, where the photon momentum is chosen as
quantization axis.

From parity conservation the following symmetry property of the small
$t$-matrix elements holds
\begin{eqnarray}\label{symmetry}
t_{-m_f -\mu -m_i}(\theta_{p},\,\theta_q,\,\phi_{pq})&=& (-1)^{-m_f+\mu+m_i}
t_{m_f \mu m_i}(\theta_{p},\,\theta_q,\,-\phi_{pq})\,.
\end{eqnarray}
Thus in contrast to single meson photoproduction on a nucleon, parity
conservation does not lead to a reduction of the number of independent
amplitudes, as has been noted already in Ref.~\cite{RoO05}. However,
this 
symmetry will allow one to classify the observables being even or odd
under the transformation $\phi_{pq}\to-\phi_{pq}$.

\section{Observables}\label{polarizObs}

In this section we briefly review the main steps for deriving all
possible observables for the present reaction as developed for
$\pi^0\eta$-photoproduction on the nucleon in Ref.~\cite{FiA11}.  It
also will allow us to introduce a more compact notation and to correct
some misprints in Ref.~\cite{FiA11}.

The basic quantity is the
following general trace with respect to the spin degrees of freedom of
photon and initial and final nucleon
\begin{equation}
A_{I'M'}= c_{kin} tr(T^\dagger\tau^{f,[I']}_{M'}e^{-iM' \phi_q}T\rho_i)\,,\label{trace}
\end{equation}
with $c_{kin}$ as a kinematical factor
\beq
c_{kin}(q,\Omega_q,\Omega_{pq})=\frac{1}{(2\pi)^5}\,
\frac{M_p^2}{E_{i}+p_i}\,\frac{1}{8\omega_\gamma\omega_q}
\frac{p_{ p}^2}{p_{p}(\omega_2+E_{f})+\frac{(\vec{q}_2+\vec{p}_f)
\cdot\vec{p}_{p}}{p_{ p}(M_p+m_2)}(E_{f}m_2-\omega_2 M_p)}\,,
\eeq
and where $\rho_i$ denotes the density matrix of the initial spin
degrees of freedom of photon and nucleon, and
$\tau^{f,[I']}_{M'}$ a spin operator with respect to the final nucleon
spin space (see Ref.~\cite{FiA11} for details). The trace has the property
\beq
A_{I'M'}^*=(-)^{M'}A_{I'-M'}\,. \label{aim-star}
\eeq
Differential cross section and recoil polarization components are
obtained from
\beq
A_{I'M'}^{\pm}=\frac{1}{2}\,(A_{I'M'}\pm (-)^{M'}A_{I'-M'})\,.\label{AIM}
\eeq
Namely, the differential cross section including
all possible polarization effects is given by
\begin{equation}
\frac{d^5\sigma}{ d^3 q d\Omega_{pq}}
= A_{00}^+\,,\label{diffcrossx}
\end{equation}
where $\Omega_{pq}=(\theta_p,\phi_q-\phi_p)$,
and the recoil nucleon polarization components with respect to the
active particle frame are given by
\beqa
P_{x}\frac{d^5\sigma}{ d^3 q d\Omega_{pq}}
&=&-\sqrt{2}\,
A_{11}^+ = -\sqrt{2}\,\mathrm{Re}\,A_{11}^+
\, ,\label{recoilpolx}\\
P_{y}\frac{d^5\sigma}{ d^3 q d\Omega_{pq}}
&=&\sqrt{2}\,i\,
A_{11}^-= -\sqrt{2}\,\mathrm{Im}\,A_{11}^-
\, ,\label{recoilpoly}\\
P_{z}\frac{d^5\sigma}{ d^3 q d\Omega_{pq}}
&=&A_{10}^+
\, .\label{recoilpolz}
\eeqa
In view of eq.~(\ref{aim-star}), the quantity $A_{I'M'}^{+}$ is real and
$A_{I'M'}^{-}$ is purely imaginary.
Obviously, $A_{I'0}^{-}$ vanishes and one has $A_{I'0}^{+}=A_{I'0}$.

Explicitly the general trace becomes
\beqa
A_{I'M'}&=&\frac{1}{2} \,\sum_{IM} P_I^p\,e^{iM
\phi_{q s}}\,d^I_{M0}(\theta_s)\, 
\sum_\mu \Big[(1+\mu P^\gamma_c)\,
u^{\mu\mu}_{I'M';IM}
-P^\gamma_\ell\,u^{\mu-\mu}_{I'M';IM}
e^{-2i\mu\phi_{q\gamma}}\Big]\,,
\eeqa
where $|P^\gamma_c|$ and $P^\gamma_\ell$ describe the degrees of
circular and linear polarization, respectively, and 
$\phi_{q\gamma}=\phi_q-\phi_\gamma$, where $\phi_\gamma$ denotes the
angle of maximal linear polarization. With respect to the nucleon
polarization parameters $P^p_I$, one has $P^p_{0}=1$, and $P^p_{1}$ describes
the degree of nucleon polarization along a direction with spherical
angles $\Omega_s=(\theta_s,\phi_s)$ and $\phi_{q
  s}=\phi_q-\phi_s$. Furthermore, we have defined
\beqa
u_{I'M';IM}^{\mu'\mu}(q,\,
\theta_q,\, \theta_{ p},\, \phi_{pq})&=& c_{kin}\,
\widehat {I'}\widehat I\,\sum_{m_f m_f' m_i m_i'}(-1)^{m_f'-m_i} \left(
\begin{matrix}
\frac{1}{2}& \frac{1}{2}&I' \cr m_f&-m_f'&M' \cr
\end{matrix} \right)
\left(
\begin{matrix}
\frac{1}{2}& \frac{1}{2}&I \cr m_i&-m_i'&-M \cr
\end{matrix} \right)\nonumber\\
&& \hspace{2cm}\times\ t^*_{m_f' \mu' m_i'}(q,\, \theta_q,\, \theta_{ p},\,
\phi_{pq}) \,t_{m_f \mu m_i}(q,\, \theta_q,\, \theta_{ p},\,
\phi_{pq})\,.\label{uim}
\eeqa
These quantities have the following symmetry properties: \\
(i) for complex conjugation one finds
\beq
\Big(u_{I'M';IM}^{\mu'\mu}(q,\, \theta_q,\, \theta_{ p},\,
\phi_{pq})\Big)^*= (-1)^{M'+M}\,u_{I'-M';I-M}^{\mu\mu'}(q,\, \theta_q,\,
\theta_{ p},\, \phi_{pq})\,,
\label{symma}
\eeq
(ii) for reversing the sign of the photon helicities $\mu$ and
$\mu^{\prime}$ from parity conservation (see eq.~(\ref{symmetry}))
\beq
u_{I'M';IM}^{-\mu'-\mu}(q,\, \theta_q,\, \theta_{ p},\, \phi_{pq})=
(-1)^{I'+M'+I+M+\mu'+\mu}\,u_{I'-M';I-M}^{\mu'\mu}
(q_,\, \theta_q,\, \theta_{p},\, -\phi_{pq})\,.
\label{symmb}
\eeq
A specific consequence of the symmetry in eq.~(\ref{symma}) is that
$u_{I'0;I0}^{\mu\mu}$ is real. Combining these two properties results
in
\beq
\Big(u_{I'M';IM}^{\mu'\mu}(q,\, \theta_q,\, \theta_{ p},\,
\phi_{pq})\Big)^*= (-1)^{I'+I+\mu'+\mu}\,
u_{I'M';IM}^{-\mu-\mu'}(q,\, \theta_q,\, \theta_{ p},
-\phi_{pq})\,.
\label{symmab}
\eeq
As one will see below, this property leads to the aforementioned
classification of the observables.

For the separation of the various types of photon polarization we introduce for
$\alpha\in\{0,c,\ell\}$ referring to unpolarized, circularly and linearly polarized
radiation, respectively, 
\beq
u^\alpha_{I'M';IM}=\sum_{\mu'\mu}\Big((\delta_{\alpha,0}+\mu\,\delta_{\alpha,c})
\,\delta_{\mu',\mu}-\delta_{\alpha,\ell}\,\delta_{\mu',-\mu}e^{-2i\mu'\phi_{q\gamma}}
\Big)u_{I'M';IM}^{\mu'\mu}\,, 
\eeq 
or in detail 
\beqa
u^0_{I'M';IM}&=&\sum_{\mu}u_{I'M';IM}^{\mu\mu}\,,\\
u^c_{I'M';IM}&=&\sum_{\mu}\mu\,u_{I'M';IM}^{\mu\mu}\,,\\
 u^\ell_{I'M';IM}&=&-\sum_{\mu}\,e^{-2i\mu\phi_{q\gamma}} \,u_{I'M';IM}^{\mu-\mu}\,.
\eeqa
These quantities have the symmetry property according to eq.~(\ref{symma})
\beq
(u^\alpha_{I'M';IM})^*=(-)^{M'+M}\,u^\alpha_{I'-M';I-M}\,,\label{symmualpha}
\eeq
which allows one to bring the trace of eq.~(\ref{trace}) into the following form
\beq
A_{I'M'}=B_{I'M'}+(-)^{M'}\,B_{I'-M'}^*
\eeq
with
\beqa
B_{I'M'}&=&\frac{1}{2}\,\sum_{I=0}^1\sum_{M=0}^I\frac{P^p_I}{1+\delta_{M0}}
\,e^{iM\phi_{qs}}\,d^I_{M0}(\theta_s)\,\sum_{\alpha\in\{0,c,\ell\}}
P_\alpha^\gamma\,u^\alpha_{I'M';IM}\,.
\eeqa
Now it is useful to introduce the following notation for
$\alpha\in\{0,c\}$ and $M\ge 0$
\beqa
\label{v0c}
 v^{0/c}_{I'M';IM}&=&\frac{1}{1+\delta_{M0}}\,u^{0/c}_{I'M';IM}\,.
\eeqa
In view of eq.~(\ref{symmualpha}) the $v^{0/c}_{I'M';IM}$ have the
symmetry property 
\beq
(v^{0/c}_{I'M';IM})^*=(-)^{M'+M}\,v^{0/c}_{I'-M';I-M}\,.\label{symmvalpha}
\eeq
Furthermore, for all $M$ we define
\beqa
 v^\ell_{I'M';IM}&=&-u^{1-1}_{I'M';IM}\,
\label{vl}
\eeqa
and use according to eq.~(\ref{symma})
\beq
u^{-11}_{I'M';IM}=-(-)^{M'+M}\,v^{\ell\,*}_{I'-M';I-M}\,.
\label{vlminus}
\eeq
With the help of the combined symmetry of eq.~(\ref{symmab}) one finds
the following behaviour under the transformation $\phi_{pq}\to -\phi_{pq}$
\beqa
v^{0/\ell}_{I'M';IM}(q_,\, \theta_q,\, \theta_{p},\, -\phi_{pq})&=&
(-)^{I'+I}(v^{0/\ell}_{I'M';IM}(q_,\, \theta_q,\, \theta_{p},\,
\phi_{pq}))^*\,,\label{parity-a}\\
v^{c}_{I'M';IM}(q_,\, \theta_q,\, \theta_{p},\, -\phi_{pq})&=&
-(-)^{I'+I}(v^{c}_{I'M';IM}(q_,\, \theta_q,\, \theta_{p},\,
\phi_{pq}))^*\,.
\label{parity-b}
\eeqa
In view of the fact that real and imaginary parts of
$v^\alpha_{I'M';IM}$ represent the observables (see below), this property allows
the classification of them into even and odd with respect to this
transformation.

Finally, one obtains
\beqa
B_{I'M'}&=&\frac{1}{2}\,\sum_{I=0}^1\sum_{M=0}^IP^p_I
\,e^{iM\phi_{qs}}\,d^I_{M0}(\theta_s)\,\Big(
\sum_{\alpha\in\{0,c\}}P_\alpha^\gamma\,v^\alpha_{I'M';IM}
\nonumber\\
&&\hspace*{2cm}+e^{-2i\phi_{q\gamma}} \,v^\ell_{I'M';IM}+(-)^{M'+M}\,
e^{2i\phi_{q\gamma}} \,v^{\ell\,*}_{I'-M';I-M}\Big)
\,.
\eeqa

For the quantities in eq.~(\ref{AIM}) one finds
\beqa
A^+_{I'M'}&=&\mathrm{Re}\,(B_{I'M'}+(-)^{M'}\,B_{I'-M'})\,,\\
A^-_{I'M'}&=&i\,\mathrm{Im}\,(B_{I'M'}+(-)^{M'}\,B_{I'-M'})\,.
\eeqa
Thus one obtains for the differential cross section and the recoil
polarization according to eqs.~(\ref{diffcrossx}) through
(\ref{recoilpolz}) 
\beqa
\frac{d^5\sigma}{ d^3 q d\Omega_{pq}}
&=&2\, \mathrm{Re}\,B_{00}\,,\label{diffx}\\
P_{x}\frac{d^5\sigma}{ d^3 q d\Omega_{pq}}
&=& -\sqrt{2}\,\mathrm{Re}\,(B_{11}-B_{1-1})
\, ,\label{polx}\\
P_{y}\frac{d^5\sigma}{ d^3 q d\Omega_{pq}}
&=&-\sqrt{2}\,\mathrm{Im}\,(B_{11}+B_{1-1})
\, ,\\
P_{z}\frac{d^5\sigma}{ d^3 q d\Omega_{pq}}
&=&2\, \mathrm{Re}\,B_{10}\label{polz}
\, .
\eeqa
Now we will proceed to list explicit expressions for the differential
cross section and the recoil polarization of the emerging nucleon
determining the various observables of this reaction.

\subsection{Differential cross section}
For convenience we introduce
\beqa
U _{IM}^{\alpha}&=&v_{00;IM}^{\alpha}
\label{UIM}
\eeqa
and separate real and imaginary parts according to
\beqa
U _{IM}^{\alpha}&=& T _{IM}^\alpha +i\,S _{IM}^\alpha \label{UT_iS}
\eeqa
for $\alpha\in\{0,c,\ell\}$. One should note that $S_{I0}^0$ and
$S_{I0}^c$ vanish according to eq.~(\ref{symmvalpha}).
In view of eqs.~(\ref{parity-a}) and (\ref{parity-b}) one finds as
symmetry property under the transformation $\phi_{pq}\to -\phi_{pq}$
\beq
U _{IM}^{\alpha}(-\phi_{pq})=(-)^{I+\delta_{\alpha,c}}\,
(U _{IM}^{\alpha}(\phi_{pq}))^*\,,
\eeq
i.e.\ $T _{IM}^\alpha$ is symmetric for $I=0$ and $\alpha\in\{0,\ell\}$,
and for $I=1$ and $\alpha=c$, and antisymmetric for $I=1$ and
$\alpha\in\{0,\ell\}$, and for $I=0$ and $\alpha=c$, whereas the
$S_{IM}^\alpha$'s have just the opposite behavior.

Then one obtains
explicitly for the differential cross section with
inclusion of beam and target polarization effects
\beqa
\frac{d^5\sigma(P^\gamma_c,P^\gamma_\ell,P^p_1)}{ d^3 q d\Omega_{pq}}
&=&
\frac{d^5\sigma_0}{ d^3 q d\Omega_{pq}}
\Big(1+P^\gamma_c\,\Sigma^c +P^\gamma_\ell\,
\Sigma^\ell (\phi_{q\gamma}) \nonumber\\
&&\hspace{2cm}+P^p_1
\,\Big[\Sigma^{p0}(\theta_s,\phi_{qs})+P^\gamma_c\,\Sigma^{pc}(\theta_s,\phi_{qs})
+P^\gamma_\ell\,\Sigma^{p\ell} (\theta_s,\phi_{qs},\phi_{q\gamma})\Big]
\Big) \,,
\eeqa
where the unpolarized differential cross section is given by
\beqa
\frac{d^5\sigma_0}{ d^3 q d\Omega_{pq}}&=&T^{0}_{00}\,.\label{T0_00}
\eeqa
Furthermore, one has
beam asymmetries for circular and linear photon polarization
\beqa
\Sigma^c\,T^{0}_{00}&=&T^{c}_{00}\,,\\
\Sigma^\ell(\phi_{q\gamma})\,T^{0}_{00}&=&
T^{\ell}_{00}\,\cos{2\phi_{q\gamma}}
+S^\ell_{00}\,\sin{2\phi_{q\gamma}}\,,
\eeqa
target asymmetry for a polarized target proton but unpolarized photons

\beqa
\Sigma^{p0}(\theta_s,\phi_{qs})\,T^{0}_{00}&=&
\cos{\theta_s} \,T^{0}_{10}
-\frac{\sin\theta_s}{\sqrt{2}}\, \Big(\cos\phi_{ qs} \,T^{0}_{11}
-\sin\phi_{qs}\,S^{0}_{11}\Big) \,,
\eeqa
and beam-target asymmetries for polarized radiation and an oriented
target
\beqa
\Sigma^{pc}(\theta_s,\phi_{qs})\,T^{0}_{00}&=&
\cos{\theta_s} \,T^{c}_{10}
-\frac{\sin\theta_s}{\sqrt{2}} \, \Big(\cos\phi_{ qs} \,T^{c}_{11}\,
-\sin\phi_{qs}\,S^{c}_{11}\Big)\,,\\
\Sigma^{p\ell} (\theta_s,\phi_{qs},\phi_{q\gamma})\,T^{0}_{00}&=&
\cos{\theta_s}\,\Big(\cos{2\phi_{q\gamma}} \,T^{\ell}_{10}
+\sin{2\phi_{q\gamma} \,S^\ell_{10}}\Big)\nonumber\\
&&
-\frac{\sin\theta_s}{\sqrt{2}} \,\Big(
\cos\phi_{qs}\Big[\cos2\phi_{q\gamma} \,(T^{\ell}_{11}-T^{\ell}_{1-1})
+\sin2\phi_{q\gamma} \,(S^{\ell}_{11}-S^{\ell}_{1-1})\Big]
\nonumber\\
&&+\sin\phi_{qs}\,
\Big[\sin2\phi_{q\gamma} \,(T^{\ell}_{11}+T^{\ell}_{1-1})
-\cos2\phi_{q\gamma} \,(S^{\ell}_{11}+S^{\ell}_{1-1})\Big]\Big)
\,. \label{Sp_l}
\eeqa
The $T^\alpha_{IM}$ and $S^\alpha_{IM}$ constitute all possible
observables of the differential cross section. For $\alpha\in\{0,c\}$
one has for each case four observables, namely $T^{0/c}_{00}, T^{0/c}_{10},
T^{0/c}_{11}$ and $S^{0/c}_{11}$, and for $\alpha=\ell$ eight ones,
$T^{\ell}_{IM}$ and $S^{\ell}_{IM}$ for $I=0,1$ and $M=-I,\dots,I$.
Altogether one finds 16 observables for the differential cross
section. Besides one unpolarized observable, the unpolarized
differential cross section, one has six single polarization and nine
double polarization observables. They can be separated by appropriate
choices of the polarization parameters and angles.

\subsection{Recoil Polarization}
Now we turn to the corresponding expressions for the recoil polarization of the
outgoing nucleon. The three components are determined by $B_{1M}$
according to eqs.~(\ref{polx}) through 
(\ref{polz}). For convenience we introduce for $\alpha\in\{0,c,\ell\}$
\beqa\label{RIMx}
R_{IM}^{x,\alpha}&=&
-\frac{1}{\sqrt{2}}\,(v^{\alpha}_{11;IM}-v^{\alpha}_{1-1;IM})\,\\
R_{IM}^{y,\alpha}&=&
\frac{i}{\sqrt{2}}\,(v^{\alpha}_{11;IM}+v^{\alpha}_{1-1;IM}) \,\\
R_{IM}^{z,\alpha}&=&v^{\alpha}_{10;IM}\,,\label{RIMz}
\eeqa
and separate into real and imaginary parts
\beq
R_{IM}^{x_i,\alpha}=P_{IM}^{x_i,\alpha}+i\,Q_{IM}^{x_i,\alpha}\,.
\eeq
One should note that the $R_{I0}^{x_i,0/c}$ are real. Furthermore for
$\alpha\in\{0,c\}$, $R_{IM}^{x_i,0/c}$ appear with $M\ge 0$ only.

As
symmetry property under the transformation $\phi_{pq}\to -\phi_{pq}$
one obtains
\beq
R_{IM}^{x_i,\alpha}(-\phi_{pq})=-(-)^{I+\delta_{\alpha,c}+\delta_{x_i,y}}
(R_{IM}^{x_i,\alpha}(\phi_{pq}))^*\,,
\eeq
which means $P_{IM}^{x_i,\alpha}$ is symmetric and
$Q_{IM}^{x_i,\alpha}$ antisymmetric for $x_i=x,z$ and either $I=1$ and
$\alpha\in\{0,\ell\}$ or  $I=0$ and
$\alpha=c$, furthermore for $x_i=y$ and either $I=0$ and
$\alpha\in\{0,\ell\}$ or  $I=1$ and
$\alpha=c$. In all other cases one has just the opposite
behavior.

For
later purpose we introduce also the spherical components $\mu=0,\pm1$
\beq
R_{IM}^{\mu,\alpha}=\delta_{\mu,0}\,R_{IM}^{z,\alpha}
-\frac{\mu}{\sqrt{2}}\,(R_{IM}^{x,\alpha}+i\,\mu\,R_{IM}^{y,\alpha})
=v^{\alpha}_{1\mu;IM}\,. \label{sphericalP}
\eeq
For $\alpha\in\{0,c\}$ one finds from the symmetry property in
eq.~(\ref{symmvalpha})
\beq
R_{IM}^{\mu,\alpha\,*}=(-)^{\mu+M}\,R_{I-M}^{-\mu,\alpha}\,,
\eeq
From which follows in particular for $M=0$
\beq
R_{I0}^{\mu,0/c\,*}=(-)^{\mu}\,R_{I0}^{-\mu,0/c}\,.
\eeq

Thus one obtains finally for the recoil polarization component
$P_{x_i}$
\beqa
P_{x_i}\frac{d^5\sigma(P^\gamma_c,P^\gamma_\ell,P^p_1)}{ d^3 q d\Omega_{pq}}&=&
\frac{d\sigma_0}{ d^3 q d\Omega_{pq}}
\Big( P^0 _{x_i}+P^\gamma_c\,P^c_{x_i} +P^\gamma_\ell\,
P^\ell_{x_i} (\phi_{q\gamma})
\nonumber\\&&\hspace{1.5cm}
+P^p_1
\,\Big[P^{p0}_{x_i}(\theta_s,\phi_{qs})+P^\gamma_c\,P^{pc}_{x_i}(\theta_s,\phi_{qs})
+P^\gamma_\ell\,P^{p\ell}_{x_i}(\theta_s,\phi_{qs},\phi_{q\gamma})\Big]
\Big)
\eeqa
with recoil polarizations for unpolarized beam and target
\beqa
P^0 _{x_i} \,T^{0}_{00}&=&P^{x_i,0}_{00}\,,
\eeqa
as well as beam asymmetries for circularly and linearly polarized
photons
\beqa
P^c_{x_i}\,T^{0}_{00}&=&P^{x_i,c}_{00}\,,\\
P^\ell_{x_i}(\phi_{q\gamma})\,T^{0}_{00}&=&
\cos{2\phi_{q\gamma}} \,P^{x_i,\ell}_{00}
+\sin{2\phi_{q\gamma}} ,Q^{x_i,\ell}_{00}\,,
\eeqa
target asymmetry for a polarized proton target
\beqa
P^{p0}_{x_i}(\theta_s,\phi_{qs}) \,T^{0}_{00}&=&
\cos{\theta_s} \,P^{x_i,0}_{10}
-\frac{\sin\theta_s}{\sqrt{2}}\,\Big(\cos\phi_{ qs}\, P^{x_i,0}_{11}
-\sin\phi_{ qs}\, Q^{x_i,0}_{11}\Big)\,,
\eeqa
and beam-target asymmetries
\beqa
P^{pc}_{x_i}(\theta_s,\phi_{qs}) \,T^{0}_{00}&=&
\cos{\theta_s} \,P^{x_i,c}_{10}
-\frac{\sin\theta_s }{\sqrt{2}}\,\Big(\cos\phi_{qs} \,P^{x_i,c}_{11}
-\sin\phi_{qs}\,Q^{x_i,c}_{11}\Big)\,,\\
P^{p\ell}_{x_i}(\theta_s,\phi_{qs},\phi_{q\gamma})\,T^{0}_{00}&=&
\cos{\theta_s}\Big(\cos{2\phi_{q\gamma}} \,P^{x_i,\ell}_{10}
+\sin{2\phi_{q\gamma} \,Q^{x_i,\ell}_{10}}\Big)\nonumber\\
&&
-\frac{\sin\theta_s}{\sqrt{2}}\,\Big(
\cos\phi_{qs}\,\Big[\cos2\phi_{q\gamma} \,(P^{x_i,\ell}_{11}-P^{x_i,\ell}_{1-1})
+\sin2\phi_{q\gamma} \,(Q^{x_i,\ell}_{11}-Q^{x_i,\ell}_{1-1})\Big]
\nonumber\\
&&\hspace{1cm}-
\sin\phi_{qs}\,\Big[
\cos2\phi_{q\gamma} (Q^{x_i, \ell}_{11}+Q^{x_i,\ell}_{1-1})-
\sin2\phi_{q\gamma} (P^{x_i, \ell}_{11}+P^{x_i,\ell}_{1-1})\Big]\Big) \,,
\,\label{recoil-pol-p}
\eeqa
constituting 48 observables for the recoil polarization components.
Of these three are single, eighteen double and twenty seven triple
polarization observables. They can be separated by appropriate choices
of the polarization parameters and angles.

Together with the 16 observables of the differential cross section this gives a total
number of 64 observables which number coincides with the maximal number of linearly
independent quadratic hermitean forms one can form from eight independent complex
amplitudes. On the other hand the eight complex amplitudes with one arbitrary phase
constitute fifteen independent parameters. Thus a minimal set of observables for the
determination of these amplitudes should comprise at least fifteen observables, a
so-called complete set. Although the above 64 observables are linearly independent there
exist quadratic relations between them, and thus it is a challenge to find minimal
(complete) sets of observables.

\subsection{Observables in terms of $t$-matrix elements}

In view of a detailed determination of the $t$-matrix elements
from observables it is useful to have explicit expressions of the
latter as linear forms in the bilinear terms
${\cal T}_{j'j}=t_{j'}^*t_j$. Here $j=\{m_f\mu\, m_i\}$ numbers the
$t$-matrix elements according to Table~\ref{tab1}.

\begin{table}[h]
\caption{Enumeration $j$ of the small $t_j$-matrix elements with
  $j=\{m_f\mu\, m_i\}$.}
\begin{ruledtabular}
\begin{tabular}{c|rrrrrrrr}
$j$ & 1 & 2 & 3 & 4 & 5 & 6 & 7 & 8\\
\colrule
$m_f$ & $1/2$ &  $-1/2$ & $1/2$ & $-1/2$  & $1/2$ &  $-1/2$ & $1/2$ & $-1/2$ \\
$\mu $ & $1$ & $1$ &  $-1$ & $-1$ & $1$ & $1$ &  $-1$ & $-1$ \\
$m_i$ & $1/2$ & $1/2$ &  $1/2$ & $1/2$ & $-1/2$ & $-1/2$ &  $-1/2$ & $-1/2$ \\
\end{tabular}
\end{ruledtabular}
\label{tab1}
\end{table}

The basic quantities in which all observables are expressed are the
$v^{\alpha}_{I'M';IM}$ as defined in eqs.~(\ref{v0c}) and (\ref{vl}) 
\beq
v^\alpha_{I'M';IM}=c_{kin}\,\sum_{j'j}\,{C}^{I'M';IM}_{j'j}\,
f^{\alpha,M}_{j'j}\,
{\cal T}_{j'j} \eeq with \beqa {C}^{I'M';IM}_{j'j}
&=&{C}^{I'M';IM}_{\{m_f'\mu 'm_i'\}\{m_f\mu\, m_i\}}\nonumber\\
&=&(-)^{m_f'-m_i}\,\widehat I'\widehat I\left(
\begin{matrix}
\frac{1}{2}& \frac{1}{2}&I' \cr m_f&-m_f'&M' \cr
\end{matrix} \right)
\left(
\begin{matrix}
\frac{1}{2}& \frac{1}{2}&I \cr m_i&-m_i'&-M \cr
\end{matrix} \right)
\,,\label{Cjj}
\eeqa
and
\beq
f^{\alpha,M}_{j'j}=f^\alpha _{\{m_f'\mu 'm_i'\}\{m_f\mu\, m_i\}}=
(\delta_{\alpha,0}+\mu \,\delta_{\alpha,c})
\frac{\delta_{\mu '\mu }}{1+\delta_{M,0}}
-\delta_{\alpha,\ell}\delta_{\mu ',1}\delta_{\mu ,-1}\,.
\eeq
Evaluation of the observables in eqs.~(\ref{UIM}) and (\ref{RIMx})
through (\ref{RIMz}) yields then the
expressions listed in appendix~\ref{obs-titj}.

\subsection{Bilinear $T$-matrix expressions in term of observables}
With respect to the question of a minimal set of observables needed
for a complete analysis we now will derive explicit expressions
for ${\cal T}_{j'j}$ in terms of observables.
The starting point is eq.~(\ref{uim}) for
$u^{\mu'\mu}_{I'M';IM}$, which are the
basic quantities for all observables in terms of the $t$-matrix
elements. It is easily inverted yielding with $j'=\{m_f' \mu' m_i'\}$
and $j=\{m_f\mu\, m_i\}$
\beq
{\cal T}_{j'j}=\frac{1}{c_{kin}}\,
\sum_{I'M'IM} {C}^{I'M';IM}_{j'j}\,u^{\mu'\mu}_{I'M';IM}
\delta_{M,m_i-m_i'}\delta_{M',m_f'-m_f}\,.
\eeq
where
${C}^{I'M';IM}_{j'j}$ is given in eq.~(\ref{Cjj}). The next step is to
express the $u^{\mu'\mu}_{I'M';IM}$ by the quantities
$v^{\alpha}_{I'M';IM}$. According to the eqs.~(\ref{v0c}), (\ref{vl})
and (\ref{vlminus}) one obtains
\beqa
u^{\mu'\mu}_{I'M';IM}&=&
\frac{1+\delta_{M,0}}{2}\,\delta_{\mu',\mu}\,
(v^{0}_{I'M';IM}+\mu\, v^{c}_{I'M';IM})
\nonumber\\&&
-\delta_{\mu',1}\,\delta_{\mu,-1}\,
v^{\ell}_{I'M';IM}-\delta_{\mu',-1}\delta_{\mu,1}(-)^{M'+M} \,
v^{\ell \, *}_{I'-M';I-M}\,.
\eeqa

As final step we relate $v^{\alpha}_{I'M';IM}$ to the various
observables of the differential cross section and the recoil
polarization components using eqs.~(\ref{UIM}) and (\ref{sphericalP})
\beqa
v^{0/c}_{I'M';IM}&=&\delta_{I',0}\,(\delta_{M,0}\,T^{0/c}_{I0}
+\delta_{M,1}\,U^{0/c}_{11}-\delta_{M,-1}\,U^{0/c\,*}_{11})\nonumber\\
&&+\delta_{I',1}\,(\delta_{M,0}\,P^{M',0/c}_{I0}
+\delta_{M,1}\,R^{M',0/c}_{11}-\delta_{M,-1}\,(-)^{M'}R^{M',0/c\,*}_{11})\,,\\
v^{\ell}_{I'M';IM}&=&\delta_{I',0}\,U^{\ell}_{IM}+\delta_{I',1}\,R^{M',\ell}_{IM}
\,.
\eeqa
Thus one obtains for ${\cal T}_{j'j}$ in terms of
observables
\beqa
{\cal T}_{j'j}&=&\frac{1}{c_{kin}}\,
\sum_{I'M'IM} {C}^{I'M';IM}_{j'j}\,
\delta_{M,m_i-m_i'}\delta_{M',m_f'-m_f}\nonumber\\
&&\times\Big\{
\frac{1+\delta_{M,0}}{2}\,\delta_{\mu',\mu}\,
\Big[\delta_{I',0}\,\Big(\delta_{M,0}\,(T^{0}_{I0}+\mu\,T^{c}_{I0})
+\delta_{M,1}\,(U^{0}_{11}+\mu\,U^{c}_{11})
-\delta_{M,-1}\,(U^{0\,*}_{11}+\mu\,U^{c\,*}_{11})\Big)\nonumber\\
&&+\delta_{I',1}\,\Big (\delta_{M,0}\,(P^{M',0}_{I0}+\mu\,P^{M',c}_{I0})
+\delta_{M,1}\,(R^{M',0}_{11}+\mu\,R^{M',c}_{11})
-\delta_{M,-1}\,(-)^{M'}(R^{M',0\,*}_{11}+\mu\,R^{M',c\,*}_{11})\Big) \Big]
\nonumber\\
&&-\delta_{\mu',1}\,\delta_{\mu,-1}\,\Big[
\delta_{I',0}\,U^{\ell}_{IM}+\delta_{I',1}\,R^{M',\ell}_{IM}\Big]
-\delta_{\mu',-1}\delta_{\mu,1}(-)^{M'+M} \,\Big[
\delta_{I',0}\,U^{\ell\,*}_{I-M}+\delta_{I',1}\,R^{-M',\ell \,*}_{I-M}\Big]\Big\}
\,.
\eeqa
A listing of the resulting expressions is given in
appendix~\ref{titj-obs}, where also a graphical representation is
introduced.

\section{On complete sets of polarization observables}
\label{completesets}
In this section we will consider two strategies for finding a minimal
complete set of observables allowing the determination of all
$t$-matrix elements up to an arbitrary phase. The first one was
developed in Ref.~\cite{ArLT98} 
and applied to the analysis of deuteron electro- and
photodisintegration in Ref.~\cite{ArLT00}. Recently, it
has also been applied to the analysis of photoproduction of two
pseudoscalar mesons on a nucleon within a truncated partial wave
approach~\cite{FiA13}. The second method was developed and applied in
Ref.~\cite{ArLT00} again to deuteron electro- and
photodisintegration.

\subsection{First Method}\label{sub-methodA}

We start with a brief description of the salient features of this
method as reported in Refs.~\cite{ArLT98,ArLT00}. The idea is as
follows: Given for an $n$-dimensional t-matrix 
a minimal set of $m=2n-1$ observables 
\beq 
{\cal
  O}^\alpha=\sum_{i,j=1,n} t_i^*H^\alpha_{ij}t_j\,,\quad
\alpha=1,\dots,m\,, 
\eeq 
constituting a set of $m$ 
hermitean quadratic forms in the $t$-matrix elements, of which
$t_{i_0}$ is chosen to be real, then a necessary condition for the
invertability is that the associated Jacobian is nonvanishing in the
vicinity of a solution.  The evaluation of the Jacobian then leads to 
the following condition: For each of the $n\times n$ matrices 
\beq
H^\alpha_{ij}=A^\alpha_{ij}+i\,B^\alpha_{ij}\,, 
\eeq 
associated with the observable ${\cal   O}^\alpha$, where
$A^\alpha_{ij}$ is a real symmetric and $B^\alpha_{ij}$ a real 
antisymmetric matrix, one constructs a $m\times m$-matrix 
\beq
\widetilde H^\alpha=\left( 
\begin{matrix}
A^\alpha & (\widetilde B^\alpha)^T \cr
\widetilde B^\alpha &\widehat A^\alpha \cr
\end{matrix} \right)\,.
\eeq 
Here, $\widehat A^\alpha$ is obtained from $A^\alpha$ by canceling the
$i_0$-th row and column, and $\widetilde B^\alpha$ from $B^\alpha$ by
canceling the $i_0$-th row. For all possible sets $\{k_1,\dots,k_m \}$
with $k_\alpha \in \{1,\dots,m\}$ one builds by choosing from
$\widetilde H^\alpha$ the $k_\alpha$-th column the matrix 
\beq
\widetilde W(k_1,\dots,k_m)=\left(\begin{array}{ccc}
\widetilde H^1_{1k_1} & \cdots &\widetilde  H^m_{1k_m} \\
\vdots & &\vdots \\ \widetilde H^1_{mk_1} & \cdots & \widetilde H^m_{mk_m}
\end{array}\right)
\,.
\eeq
One should note that the $k_\alpha$ need not be different. Now the
condition is that at least one of the determinants of $\widetilde
W(k_1,\dots,k_m)$ is nonvanishing. This condition, however, is in
general not sufficient in case that several of these determinants are
nonvanishing. If only one determinant is nonvanishing, than this
condition is also sufficient. Moreover, one might encounter quadratic
ambiguities in the solution.

Turning now to the present reaction, one readily notes that, according
to the explicit listing of all observables in terms of the $t$-matrix
elements, all
matrices $H^\alpha_{ij}$ have a simple structure. They are either real
symmetric, i.e.\ of type $A^\alpha$, or imaginary antisymmetric, i.e.\
of type $iB^\alpha$. Moreover, they have for each row and each column
at most only one nonvanishing entry. Thus the associated matrices
$\widetilde H^\alpha$ are easily constructed and have a similar
structure. For this reason, it turns out, that for any selection of 15
observables the above criterion is fulfilled. However, in all cases
one can find more than one non-vanishing determinant.

As an example, we have selected the following set of fifteen
observables, guided by their representation in terms of ${\cal
  T}_{j'j}$ (see appendix~\ref{titj-obs}): $T^0_{0 0}$, $U^0_{1
  1}=T^0_{1 1}+i\,S^0_{1 1}$, $U^c_{1 1}=T^c_{1 1}+i\,S^c_{1 1}$,
$R^{z,0}_{1 1}=P^{z,0}_{1 1}+i\,Q^{z,0}_{1 1}$, $R^{z,c}_{1
  1}=P^{z,c}_{1 1}+i\,Q^{z,c}_{1 1}$, $U^\ell_{0 0}=T^\ell_{0
  0}+i\,S^\ell_{0 0}$, $U^\ell_{1 0}=T^\ell_{1 0}+i\,S^\ell_{1 0}$,
$P^{x+iy,0}_{0 0}=P^{x,0}_{0 0}+i\,P^{y,0}_{0 0}$. This set contains
one unpolarized observable, four single, eight double and two triple
polarization observables.

In
appendix~\ref{methodA} it is shown, how all seven matrix elements
$t_2,\dots,t_8$ can be expressed by $t_1$ and the chosen observables.
In detail one finds $t_j=\frac{\sigma_j}{t_1^*}$ for $j=2,3,5,8$
with
\beq
\sigma_2=c_{12}\,,\quad \sigma_3=c_{13}\,,\quad
\sigma_5=c_{51}^* \,,\quad \sigma_8=\frac{c_{12}\,c_{84}^*}{c_{24}^*}\,,
\eeq
and $t_j=\tau_j\,t_1$ for $ j=4,6,7$ with
\beq
\tau_4=\frac {c_{24}}{c_{12}^*}\,,\quad
\tau_6=\frac {c_{62}^*}{c_{12}^*}\,,\quad
\tau_7=\frac {c_{73}^*}{c_{13}^*}\,.
\eeq
The various complex constants $c_{ij}$, expressed in terms of
observables, may be found in the appendix~\ref{methodA}. The constants
$c_{12}$ and $c_{13}$ contain quadratic ambiguities.

Finally, the remaining matrix element $t_1$ is
obtained from the unpolarized differential cross section
in eq.~(\ref{t00}), i.e.\
\beqa
c_0&=&4\,c_{kin} \,T^0_{0 0}=
a_1\,|t_1|^2+\frac{b_1}{|t_1|^2}\,,
\eeqa
with
\beq
a_1=1+|\tau_4|^2+|\tau_6|^2+|\tau_8|^2\,,\quad
b_1=|\sigma_2|^2+|\sigma_3|^2
+|\sigma_5|^2+\sigma_8|^2\,.
\eeq
It has as solution
\beqa
|t_1|^2&=&\frac{1}{2a_1}\Big(c_0\pm
\sqrt{(c_0)^2-4\,a_1\,b_1}\Big)\,,\label{solutiont1}
\eeqa
introducing a third ambiguity. However, it turns out, that some of
these ambiguities are eliminated by the condition
\beq
(c_0)^2-4\,a_1\,b_1\ge 0\,.\label{condition}
\eeq
Indeed, taking a specific numerical example, we found, that only the 
ambiguity of eq.~(\ref{solutiont1}) remains, which is easily resolved
by selecting one additional 
observables, for example, $P^{z,0}_{00}$. Altogether, these sixteen
observables allow one to determine uniquely the eight complex
$t$-matrix elements.

\subsection{Second Method}\label{sub-methodB}

Another possibility of constructing a complete set of polarization
observables is to study first the representation of the bilinear
$t$-matrix products ${\cal T}_{j'j}$ in terms of observables. In
appendix~\ref{titj-obs} we have listed explicit expressions and also
outlined a graphical representation as devised in
Ref.~\cite{ArLT00}. It turns out that they can be divided
into groups according to the participating observables. This division
is unique and there is no overlap of observables between different
groups.  Altogether one obtains eight groups, one containing the eight
diagonal terms $|t_j|^2$ with eight observables, and seven groups for
the 28 interference terms, each containing four interference terms
with eight observables.

One can now try to combine the various interference terms in a
complete chain of interference terms $t_{j_1 j_2},\dots,t_{j_{n-1} j_n}$
with $(j_1,\dots,j_n)$ as a permutation of $(1,\dots,n)$. The ideal case is,
that the participating observables of such a chain plus an
additional independent observable constitute a minimal complete set,
i.e.\ are sufficient for the determination 
of all $t$-matrix elements. However, such an ideal situation is seldom
found. In fact, for the present reaction this is not the case as the
graphical representations of the various groups in
appendix~\ref{titj-obs} demonstrate. However, we can utilize these
representations by combining various groups in order to construct such
a chain. In such combinations one finds closed loops which constitute
higher order relations between observables, which then can be used for
the elimination of superfluous observables. 

\begin{figure}[h]
\begin{center}
\includegraphics[scale=.4]{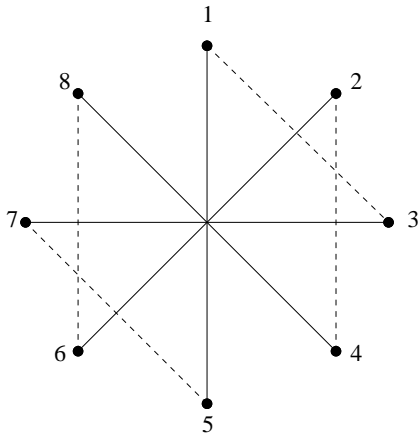}
\end{center}
\caption{Combined graphical representation of the groups ``A1''
  (solid) and   ``B'' (dashed).}
\label{figA1-B}
\end{figure}

For example, considering the observables of the
differential cross section and recoil polarization component $P_z$ and
combining group ``A1'', containing $U^0_{11}$, $U^c_{11}$, $ R^{
  z,0}_{11}$, and $ R^{ z,c}_{11}$, with group ``B'', containing $U^{\ell}_{ 0 0}$,
$ U^{\ell}_{ 1 0}$, $ R^{ z,\ell}_{ 0 0}$, and $ R^{ z,\ell}_{ 1 0}$,
one obtains the pattern displayed in
Fig.~\ref{figA1-B}. Here one can distinguish two connected groups: (i)
group ``I''
with the matrix elements $t_1$, $t_3$, $t_5$, and $t_7$, and (ii)
group ``II'' with the matrix elements of even number $t_2$, $t_4$,
$t_6$, and $t_8$. For 
each group the matrix elements  are connected by interference terms
building two four-point closed loops, namely ``1-3-7-5-1'' and
``2-4-8-6-2''.

Thus for both groups all $t$-matrix elements terms can be expressed
relative to one matrix element, for example in the first group ``I'' with
respect to $t_1$, i.e.
\beq
t_3=\frac{{\cal T}_{13}}{t_1^*}\,,\quad
t_5=\frac{{\cal T}_{15}}{t_1^*}\,,\quad
t_7=\frac{{\cal T}_{57}}{{\cal T}_{51}}\,t_1
=\frac{{\cal T}_{37}}{{\cal T}_{31}}\,t_1\,.
\eeq
The last equation yields because of the closed loop ``1-3-7-5-1'' a
quadratic relation between observables
\beqa
{\cal T}_{31}{\cal T}_{57}&=&{\cal T}_{37}{\cal T}_{51}\,,\label{quadrela}
\eeqa
or explicitly in terms of observables
\beqa
{2}\, (U^{\ell}_{0 0}+U^{\ell}_{1 0}+R^{z,\ell}_{0 0}+R^{z,\ell}_{1 0}) ^*
(U^{\ell}_{0 0}-U^{\ell}_{1 0}+R^{z,\ell}_{0 0}-R^{z,\ell}_{1 0})
&=&
(U^{0}_{1 1}-U^{c}_{1 1}+R^{z,0}_{1 1}-R^{z,c}_{1 1}) ^* \nonumber\\
&&\times
(U^{0}_{1 1}+U^{c}_{1 1}+R^{z,0}_{1 1}+R^{z,c}_{1 1})\,.\label{quadrela1}
\eeqa
Similarly one can express all matrix elements of the second group
``II'' in terms of say $t_2$ according to
\beq
t_4=\frac{{\cal T}_{24}}{t_2^*}\,,\quad
t_6=\frac{{\cal T}_{26}}{t_2^*}\,,\quad
t_8=\frac{{\cal T}_{48}}{{\cal T}_{42}}\,t_2
=\frac{{\cal T}_{68}}{{\cal T}_{62}}\,t_2\,.
\eeq
Again, one finds from the last equation a quadratic relation from  the
second closed loop ``2-4-8-6-2''
\beqa
{\cal T}_{42}{\cal T}_{68}&=&{\cal T}_{48}{\cal T}_{62}\,,\label{quadrelb}
\eeqa
and in terms of observables
\beqa
{2}\, (U^{\ell}_{0 0}+U^{\ell}_{1 0}-R^{z,\ell}_{0 0}-R^{z,\ell}_{1 0})^*
(U^{\ell}_{0 0}-U^{\ell}_{1 0}-R^{z,\ell}_{0 0}+R^{z,\ell}_{1 0})
&=&
(U^{0}_{1 1}-U^{c}_{1 1}-R^{z,0}_{1 1}+R^{z,c}_{1 1}) ^* \nonumber\\
&&\times
 (U^{0}_{1 1}+U^{c}_{1 1}-R^{z,0}_{1 1}-R^{z,c}_{1 1})
\,. \label{quadrelb1}
\eeqa
Formally this relation can be obtained from eq.~(\ref{quadrela1}) by the
substitutions $R^{z,\ell}_{0 0/10}\to -R^{z,\ell}_{0 0/10}$ and
$R^{z,0/c}_{11}\to -R^{z,0/c}_{11}$. These two relations can be utilized for the
elimination of the four triple polarization observables contained
in $R^{z,c}_{1 1}$ and $R^{z,\ell}_{1  0}$. This is shown in
appendix~\ref{methodB}. The remaining group contains only single and
double polarization observables.

Thus, the matrix elements  with odd numbers $t_3$, $t_5$, and $t_7$
can be expressed by the observables of ``A1'' and ``B'' and $t_1$,
while the ones with even numbers $t_4$, $t_6$, and $t_8$ can be
expressed by the same observables and
$t_2$. Of the sixteen observables of ``A1'' and ``B'' four, namely
$R^{z,c}_{11}$ and $R^{z,\ell}_{1 0}$, are eliminated, leaving
twelve observables.

\begin{figure}[h]
\begin{center}
\includegraphics[scale=.4]{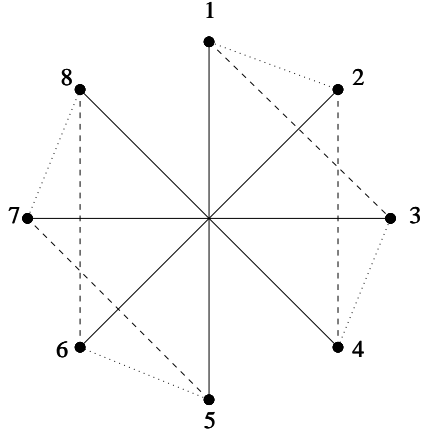}
\end{center}
\caption{Combined graphical representation of the groups ``A1''
  (solid),   ``B'' (dashed) and ``C'' (dotted).}
\label{figA1-B-C}
\end{figure}

Obviously, for a complete determination  one needs an
interference term connecting these two groups, i.e.\ an interference
term ${\cal T}_{j'j}$ with $j'$ even and $j$ odd or vice versa. Since
the interference terms given in terms of observables of the
differential cross section and the recoil polarization component $P_z$
involve $t$-matrix elements of either both even or both odd numbers
one has to choose
one of the groups of interference terms involving observables of the
recoil polarization components $P_x$ and $P_y$, i.e.\ one of the
groups ``C'' through ``D1''. For example, choosing ${\cal  T}_{12}$ as
the missing link, one has to add the group ``C'', containing
$P^{1,0}_{00}$, $ P^{ 1,c}_{00}$, $ P^{ 1,0}_{10}$, and $
P^{1,c}_{10}$, two single, four double and two triple polarization
observables. The resulting pattern is shown in Fig.~\ref{figA1-B-C}.
Now one can express $t_2$ in terms of observables and $t_1$
\beqa
t_2&=&\frac{{\cal T}_{12}}{t_1^*}\,.
\eeqa

However, now we have more observables than needed, namely twenty,
which means that six of them are superfluous and that more interrelations
must exist. In fact, adding the group ``C'' generates four more four-point
loops, namely ``1-2-4-3-1'', ``5-6-8-7-5'', ``1-2-6-5-1'', and
``3-4-8-7-3''. However, only two of the additional quadratic relations
are independent. This can be seen as follows. The two new four-point
loops ``1-2-4-3-1'' and ``5-6-8-7-5'' between the groups ``B'' and
``C'' generate as quadratic relations
\beqa
{\cal T}_{21}{\cal T}_{34}&=&{\cal T}_{24}{\cal T}_{31}\,,\label{quadrelc}\\
{\cal T}_{65}{\cal T}_{78}&=&{\cal T}_{68}{\cal T}_{75}\,.\label{quadreld}
\eeqa
and the other two loops ``1-2-6-5-1'' and ``3-4-8-7-3'' between the
groups ``A1'' and ``C'' generate
\beqa
{\cal T}_{65}{\cal T}_{21}&=&{\cal T}_{61}{\cal T}_{25}\,,\label{quadrele}\\
{\cal T}_{78}{\cal T}_{34}&=&{\cal T}_{74}{\cal
  T}_{38}\,.\label{quadrelf} 
\eeqa 
But the latter two are not independent from the previous quadratic
relations in eqs.~(\ref{quadrelc}) and (\ref{quadreld}). For example,
using eqs.~(\ref{quadrelc}) and (\ref{quadreld}) one finds 
\beq 
{\cal T}_{34}=\frac{{\cal T}_{31}{\cal T}_{24}}{{\cal T}_{21}} 
\quad\mathrm{and} \quad 
{\cal T}_{78}=\frac{{\cal T}_{75}{\cal T}_{68}}{{\cal T}_{65}}\,. 
\eeq 
Inserting these expressions into eq.~(\ref{quadrelf}) one obtains
consecutively 
\beq 
\frac{{\cal T}_{75}{\cal T}_{68}}{{\cal T}_{65}}\,
\frac{{\cal T}_{31}{\cal T}_{24}}{{\cal T}_{21}}={\cal T}_{74}{\cal
  T}_{38} 
\eeq
and thus
\beq 
{\cal T}_{31}{\cal T}_{24}= \frac{{\cal T}_{74}{\cal T}_{38}{\cal T}_{65}}
{{\cal T}_{75}{\cal T}_{68}}\,{\cal T}_{21}
=\frac{{\cal T}_{74}{\cal T}_{35}}{{\cal T}_{75}}\,{\cal T}_{21}
={\cal T}_{34}{\cal T}_{21}\,,
\eeq
which is the relation in eq.~(\ref{quadrelc})

The quadratic relations in eqs.~(\ref{quadrelc}) and (\ref{quadreld})
read in terms of observables of groups ``B'' and ``C''
\beqa
2\,(P^{ 1,0}_{00}+P^{ 1,c}_{00}+P^{ 1,0}_{10}+P^{ 1,c}_{10})^*
(P^{ 1,0}_{00}-P^{ 1,c}_{00}+P^{ 1,0}_{10}-P^{ 1,c}_{10})
&=&
(U^{\ell}_{0 0}+U^{\ell}_{1 0}+R^{z,\ell}_{0 0}+R^{z,\ell}_{1 0})^* \nonumber\\
&&\times
(U^{\ell}_{0 0}+U^{\ell}_{1 0}-R^{z,\ell}_{0 0}-R^{z,\ell}_{1 0})
\,,\label{quadrelc1}\\
2\,(P^{ 1,0}_{00}+P^{ 1,c}_{00}-P^{ 1,0}_{10}-P^{ 1,c}_{10})^*
(P^{ 1,0}_{00}-P^{ 1,c}_{00}-P^{ 1,0}_{10}+P^{ 1,c}_{10})
&=&
(U^{\ell}_{0 0}-U^{\ell}_{1 0}+R^{z,\ell}_{0 0}-R^{z,\ell}_{1 0})^* \nonumber\\
&&\times
(U^{\ell}_{0 0}-U^{\ell}_{1 0}-R^{z,\ell}_{0 0}+R^{z,\ell}_{1 0})
\,. \label{quadreld1}
\eeqa
These two complex relations would allow one to eliminate only four of
the six observables. 

However, besides the four-point loops one finds sixteen six-point
loops of which however only one is independent, which one can show
easily in the same manner as before for the four-point loops. Thus one
has one additional relation of third order between observables of all three groups. 
Taking the six-point loop ``1-2-6-5-7-3-1'', one obtains the
relation 
\beq 
{\cal
T}_{37}{\cal T}_{56}{\cal T}_{21}= {\cal T}_{26}{\cal T}_{57}{\cal
T}_{31}\,,\label{quadrelg} 
\eeq 
which reads in terms of observables 
\beqa
2\,\Big(U^0_{11}-U^c_{11}+R^{ z,0}_{11}-R^{ z,c}_{11}\Big)^* 
\Big(P^{ 1,0}_{00}+P^{1,c}_{00}-P^{ 1,0}_{10}-P^{ 1,c}_{10}
\Big) &&\nonumber\\
\times\Big(P^{ 1,0}_{00}+P^{ 1,c}_{00}+P^{ 1,0}_{10}+P^{ 1,c}_{10}
          \Big)^*
&=&
\Big(U^0_{11}+U^c_{11}-R^{ z,0}_{11}-R^{ z,c}_{11}
          \Big)^*\\
&&\times\Big(U^{\ell}_{ 0 0}-U^{\ell}_{ 1 0}+R^{ z,\ell}_{ 0 0}-R^{ z,\ell}_{ 1 0}
          \Big)
\Big(U^{\ell}_{ 0 0}+U^{\ell}_{ 1 0}+R^{ z,\ell}_{ 0 0}+R^{ z,\ell}_{ 1 0}
          \Big)^*\nonumber
\eeqa
This relation now provides the means to eliminate two more
observables. As the six observables to be eliminated we have chosen
from the Group ``C''  $P^{ 1,c}_{00}$, $P^{ 1,0}_{10}$,
and $P^{ 1,c}_{10}$. How this is done is outlined in appendix~\ref{methodB}.

Thus all seven matrix elements $t_2,\dots,t_8$ are given
by $t_1$ and fourteen observables, since the 24
observables of the groups ``A1'', ``B'', and ``C'' are reduced by five
complex relations to 14, namely
$U^0_{11}$, $U^c_{11}$, $ R^{z,0}_{11}$, $U^{\ell}_{ 0 0}$,
$ U^{\ell}_{ 1 0}$, $R^{z,\ell}_{00}$, and  $ P^{ 1,0}_{00}$. For the
determination of $t_1$ one can use again the unpolarized differential
cross section. 

Altogether we can obtain all eight $t$-matrix elements from 15
observables up to some quadratic ambiguities without the need of a
triple polarization observable. Like in the first method, two
ambiguities are ruled out by the condition in eq.~(\ref{condition}) as
we have checked by a numerical example.

\section{Conclusion}\label{conclusion}
We have presented two methods for  allowing one to choose a minimal
set of observables, which may be used for a complete determination of the
$t$-matrix elements for photoproduction of two pseudoscalar mesons on
a nucleon. The methods are based on the inversion of the exact  
expressions for all observables as Hermitean forms in $t^*_it_j$ of
the $t$-matrix elements.
We also have demonstrated that one can choose a  complete set of
observables 
without the need of triple polarization observables. This important
theoretical result reduces to some extent the pessimism around 
the realization of a complete experiment for photoproduction of two
pseudoscalars in view of a possible need of triple polarization
observables, which constitutes quite a severe condition for such an
experiment. On the other hand we are aware that, at least at present
times, our results are primarily of  theoretical interest, and still
many experimental efforts have to be undertaken towards the
achievement of conditions which will allow a practical realization of
the methods developed in the present work.

\section*{Acknowledgment}
This work was supported by the Deutsche Forschungsgemeinschaft (SFB 1044). A.F.
acknowledges additional support by the Dynasty Foundation.


\appendix
\renewcommand{\theequation}{A\arabic{equation}}
\setcounter{equation}{0}
\section{Listing of observables in terms of $t$-matrix elements}\label{obs-titj}
In this appendix we list all observables as bilinear
forms $t_{j'}^*t_j$ of the small $t$-matrix elements, where we have
introduced the notation ${\cal T}_{j'j}=t_{j'}^*t_{j}$: 

(i) Differential cross section without and with target polarization
for\\
(a) unpolarized photons:\\ 
\beqa
T^0_{0 0}&=&
\frac{c_{kin}}{4}\,(
  {\cal T}_{11}
  +{\cal T}_{22}
  +{\cal T}_{33}
  +{\cal T}_{44}
  +{\cal T}_{55}
  +{\cal T}_{66}
  +{\cal T}_{77}
  +{\cal T}_{88}
)
\,,\\
T^0_{1 0}&=&
\frac{c_{kin}}{4}\,(
  {\cal T}_{11}
  +{\cal T}_{22}
  +{\cal T}_{33}
  +{\cal T}_{44}
  -{\cal T}_{55}
  -{\cal T}_{66}
  -{\cal T}_{77}
  -{\cal T}_{88}
)
\,,\\
U^0_{1 1}&=&T^0_{1 1}+i\,S^0_{1 1}=
-\frac{c_{kin}}{\sqrt{2}}\,(
  {\cal T}_{51}
  +{\cal T}_{62}
  +{\cal T}_{73}
  +{\cal T}_{84}
)\,,
\eeqa
(b) circularly polarized photons:
\beqa
T^c_{0 0}&=&
\frac{c_{kin}}{4}\,(
  {\cal T}_{11}
  +{\cal T}_{22}
  -{\cal T}_{33}
  -{\cal T}_{44}
  +{\cal T}_{55}
  +{\cal T}_{66}
  -{\cal T}_{77}
  -{\cal T}_{88}
)
\,,\\
T^c_{1 0}&=&
\frac{c_{kin}}{4}\,(
  {\cal T}_{11}
  +{\cal T}_{22}
  -{\cal T}_{33}
  -{\cal T}_{44}
  -{\cal T}_{55}
  -{\cal T}_{66}
  +{\cal T}_{77}
  +{\cal T}_{88}
)
\,,\\
U^c_{1 1}&=&T^c_{1 1}+i\,S^c_{1 1}=
-\frac{c_{kin}}{\sqrt{2}}\,(
  {\cal T}_{51}
  +{\cal T}_{62}
  -{\cal T}_{73}
  -{\cal T}_{84}
) \,,
\eeqa
(c) and linearly polarized photons:
\beqa
U^\ell_{0 0}&=&T^\ell_{0 0}+i\,S^\ell_{0 0}=
-\frac{c_{kin}}{2}\,(
  {\cal T}_{13}
  +{\cal T}_{24}
  +{\cal T}_{57}
  +{\cal T}_{68}
)
\,,\\
U^\ell_{1-1}&=&T^\ell_{1-1}+i\,S^\ell_{1-1}=
-\frac{c_{kin}}{\sqrt{2}}\,(
  {\cal T}_{17}+{\cal T}_{28})
\,,\\
U^\ell_{1 0}&=&T^\ell_{1 0}+i\,S^\ell_{1 0}=
-\frac{c_{kin}}{2}\,(
  {\cal T}_{13}
  +{\cal T}_{24}
  -{\cal T}_{57}
  -{\cal T}_{68}
)
\,,\\
U^\ell_{1 1}&=&T^\ell_{1 1}+i\,S^\ell_{1 1}=
\frac{c_{kin}}{\sqrt{2}}\,(
  {\cal T}_{53}
  +{\cal T}_{64}
) \,.
\eeqa
(ii) Recoil polarization $P^z$ without and with target polarization for\\(a) unpolarized photons:\\
\beqa
P^{z,0}_{0 0}&=&
\frac{c_{kin}}{4}\,(
  {\cal T}_{11}
  -{\cal T}_{22}
  +{\cal T}_{33}
  -{\cal T}_{44}
  +{\cal T}_{55}
  -{\cal T}_{66}
  +{\cal T}_{77}
  -{\cal T}_{88}
)
\,,\\
P^{z,0}_{1 0}&=&
\frac{c_{kin}}{4}\,(
  {\cal T}_{11}
  -{\cal T}_{22}
  +{\cal T}_{33}
  -{\cal T}_{44}
  -{\cal T}_{55}
  +{\cal T}_{66}
  -{\cal T}_{77}
  +{\cal T}_{88}
)
\,,\\
R^{z,0}_{1 1}&=&P^{z,0}_{1 1}+i\,Q^{z,0}_{1 1}=
-\frac{c_{kin}}{\sqrt{2}}\,(
  {\cal T}_{51}
  -{\cal T}_{62}
  +{\cal T}_{73}
  -{\cal T}_{84}
) \,,
\eeqa
(b) circularly polarized photons:
\beqa
P^{z,c}_{0 0}&=&
\frac{c_{kin}}{4}\,(
  {\cal T}_{11}
  -{\cal T}_{22}
  -{\cal T}_{33}
  +{\cal T}_{44}
  +{\cal T}_{55}
  -{\cal T}_{66}
  -{\cal T}_{77}
  +{\cal T}_{88}
)
\,,\\
P^{z,c}_{1 0}&=&
\frac{c_{kin}}{4}\,(
  {\cal T}_{11}
  -{\cal T}_{22}
  -{\cal T}_{33}
  +{\cal T}_{44}
  -{\cal T}_{55}
  +{\cal T}_{66}
  +{\cal T}_{77}
  -{\cal T}_{88}
)
\,,\\
R^{z,c}_{1 1}&=&P^{z,c}_{1 1}+i\,Q^{z,c}_{1 1}=
-\frac{c_{kin}}{\sqrt{2}}\,(
  {\cal T}_{51}
  -{\cal T}_{62}
  -{\cal T}_{73}
  +{\cal T}_{84}
) \,,
\eeqa
(c) and linearly polarized photons:
\beqa
R^{z,\ell}_{0 0}&=&P^{z,\ell}_{0 0}+i\,Q^{z,\ell}_{0 0}=
-\frac{c_{kin}}{2}\,(
  {\cal T}_{13}
  -{\cal T}_{24}
  +{\cal T}_{57}
  -{\cal T}_{68}
)
\,,\\
R^{z,\ell}_{1-1}&=&P^{z,\ell}_{1-1}+i\,Q^{z,\ell}_{1-1}=
-\frac{c_{kin}}{\sqrt{2}}\,(
 {\cal T}_{17}-{\cal T}_{28}
)
\,,\\
R^{z,\ell}_{1 0}&=&P^{z,\ell}_{1 0}+i\,Q^{z,\ell}_{1 0}=
-\frac{c_{kin}}{2}\,(
  {\cal T}_{13}
  -{\cal T}_{24}
  -{\cal T}_{57}
  +{\cal T}_{68}
)
\,,\\
R^{z,\ell}_{1 1}&=&P^{z,\ell}_{1 1}+i\,Q^{z,\ell}_{1 1}=
\frac{c_{kin}}{\sqrt{2}}\,(
  {\cal T}_{53}
  -{\cal T}_{64}
) \,.
\eeqa
(iii) Recoil polarization $P^x$ and $P^y$ without and with target polarization for\\(a) unpolarized photons:
\beqa
P^{x,0}_{0 0}&=&
\frac{c_{kin}}{2}\,\mathrm{Re}\Big(
  {\cal T}_{21}
  +{\cal T}_{43}
  +{\cal T}_{65}
  +{\cal T}_{87}
\Big) \,,\\
P^{y,0}_{0 0}&=&
\frac{c_{kin}}{2}\,\mathrm{Im}\Big(
  {\cal T}_{21}
  +{\cal T}_{43}
  +{\cal T}_{65}
  +{\cal T}_{87}
\Big) \,,\\
P^{x,0}_{1 0}&=&
\frac{c_{kin}}{2}\,\mathrm{Re}\Big(
  {\cal T}_{21}
  +{\cal T}_{43}
  -{\cal T}_{65}
  -{\cal T}_{87}
\Big) \,,\\
P^{y,0}_{1 0}&=&
\frac{c_{kin}}{2}\,\mathrm{Im}\Big(
  {\cal T}_{21}
  +{\cal T}_{43}
  -{\cal T}_{65}
  -{\cal T}_{87}
\Big) \,,\\
R^{x,0}_{1 1}&=&P^{x,0}_{1 1}+i\,Q^{x,0}_{1 1}=
-\frac{c_{kin}}{\sqrt{2}}\,\Big(
  {\cal T}_{52}
  +{\cal T}_{61}
  +{\cal T}_{74}
  +{\cal T}_{83}
\Big) \,,\\
R^{y,0}_{1 1}&=&P^{y,0}_{1 1}+i\,Q^{y,0}_{1 1}=
i \,\frac{c_{kin}}{\sqrt{2}}\,\Big(
  {\cal T}_{52}
  -{\cal T}_{61}
  -{\cal T}_{74}
  +{\cal T}_{83}
\Big) \,,
\eeqa
(b) circularly polarized photons:
\beqa
P^{x,c}_{0 0}&=&
\frac{c_{kin}}{2}\,\mathrm{Re}\Big(
  {\cal T}_{21}
  -{\cal T}_{43}
  +{\cal T}_{65}
  -{\cal T}_{87}
\Big) \,,\\
P^{y,c}_{0 0}&=&
  \frac{c_{kin}}{2}\,\mathrm{Im}\Big(
  {\cal T}_{21}
  -{\cal T}_{43}
  +{\cal T}_{65}
  -{\cal T}_{87}
\Big) \,,\\
P^{x,c}_{1 0}&=&
\frac{c_{kin}}{2}\,\mathrm{Re}\Big(
  {\cal T}_{21}
  -{\cal T}_{43}
  -{\cal T}_{65}
  +{\cal T}_{87}
\Big) \,,\\
P^{y,c}_{1 0}&=&
  \frac{c_{kin}}{2}\,\mathrm{Im}\Big(
  {\cal T}_{21}
  -{\cal T}_{43}
  -{\cal T}_{65}
  +{\cal T}_{87}
\Big) \,,\\
R^{x,c}_{1 1}&=&P^{x,c}_{1 1}+i\,Q^{x,c}_{1 1}=
-\frac{c_{kin}}{\sqrt{2}}\,\Big(
  {\cal T}_{52}
  +{\cal T}_{61}
  -{\cal T}_{74}
  -{\cal T}_{83}
\Big) \,,\\
R^{y,c}_{1 1}&=&P^{y,c}_{1 1}+i\,Q^{y,c}_{1 1}=
i\,\frac{c_{kin}}{\sqrt{2}}\,\Big(
  {\cal T}_{52}
  -{\cal T}_{61}
  +{\cal T}_{74}
  -{\cal T}_{83}
\Big) \,,
\eeqa
(c) and linearly polarized photons:
\beqa
R^{x,\ell}_{0 0}&=&P^{x,\ell}_{0 0}+i\,Q^{x,\ell}_{0 0}=
-\frac{c_{kin}}{2}\,\Big(
  {\cal T}_{14}
  +{\cal T}_{23}
  +{\cal T}_{58}
  +{\cal T}_{67}
\Big) \,,\\
R^{y,\ell}_{0 0}&=&P^{y,\ell}_{0 0}+i\,Q^{y,\ell}_{0 0}=
i\frac{c_{kin}}{2}\,\Big(
  {\cal T}_{14}
  -{\cal T}_{23}
  +{\cal T}_{58}
  -{\cal T}_{67}
\Big) \,,\\
R^{x,\ell}_{1-1}&=&P^{x,\ell}_{1-1}+i\,Q^{x,\ell}_{1-1}=
-\frac{c_{kin}}{\sqrt{2}}\,\Big(
  {\cal T}_{18}
  +{\cal T}_{27}
\Big) \,,\\
R^{y,\ell}_{1-1}&=&P^{y,\ell}_{1-1}+i\,Q^{y,\ell}_{1-1}=
i\,\frac{c_{kin}}{\sqrt{2}}\,\Big(
  {\cal T}_{18}
  -{\cal T}_{27}
\Big) \,,\\
R^{x,\ell}_{1 0}&=&P^{x,\ell}_{1 0}+i\,Q^{x,\ell}_{1 0}=
-\frac{c_{kin}}{\sqrt{2}}\,\Big(
  {\cal T}_{14}
  +{\cal T}_{23}
  -{\cal T}_{58}
  -{\cal T}_{67}
\Big) \,,\\
R^{y,\ell}_{1 0}&=&P^{y,\ell}_{1 0}+i\,Q^{y,\ell}_{1 0}=
i\,\frac{c_{kin}}{\sqrt{2}}\,\Big(
  {\cal T}_{14}
  -{\cal T}_{23}
  -{\cal T}_{58}
  +{\cal T}_{67}
\Big) \,,\\
R^{x,\ell}_{1 1}&=&P^{x,\ell}_{1 1}+i\,Q^{x,\ell}_{1 1}=
\frac{c_{kin}}{\sqrt{2}}\,\Big(
  {\cal T}_{54}
  +{\cal T}_{63}
\Big) \,,\\
R^{y,\ell}_{1 1}&=&P^{y,\ell}_{1 1}+i\,Q^{y,\ell}_{1 1}=
-i\,\frac{c_{kin}}{\sqrt{2}}\,\Big(
  {\cal T}_{54}
  -{\cal T}_{63}
\Big) \,.
\eeqa

\renewcommand{\theequation}{B\arabic{equation}}
\setcounter{equation}{0}
\section{Listing of the bilinear $t$-matrix expressions in terms of observables}
\label{titj-obs}
In this appendix we list explicit expressions of the bilinear forms
${\cal T}_{j'j}=t_{j'}^* t_j$ in terms of observables.
We have divided them into groups according to the
type of participating observables. Each group is accompanied by a
graphical representation as originally devised in Ref.~\cite{ArLT00}
in which each 
matrix element $t_j$ is represented by a point labeled ``$j$'' on a
circle, and to a bilinear term $t_it_j$ is associated a straight
line connecting the points ``$i$" and ``$j$''.
As pointed out in Ref.~\cite{ArLT00} a closed loop with four points leads
to a quadratic relation between observables because of the following,
immediately evident property
\beq
{\cal T}_{ab}{\cal T}_{cd}={\cal T}_{ad}{\cal T}_{cb}
\,.\label{quadrel}
\eeq
Two special cases follow from this relation
\beqa
{\cal T}_{aa}{\cal T}_{bc} &=&{\cal T}_{ac}{\cal T}_{ba}
\,,\label{quadrel3}\\
{\cal T}_{aa}{\cal T}_{bb} &=&|{\cal T}_{ab}|^2
\,. \label{quadrel2}
\eeqa
Though these relations are trivial in terms of $t$-matrix elements,
they are not if expressed in terms of observables.

\begin{figure}[h]
\begin{center}
\includegraphics[scale=.4]{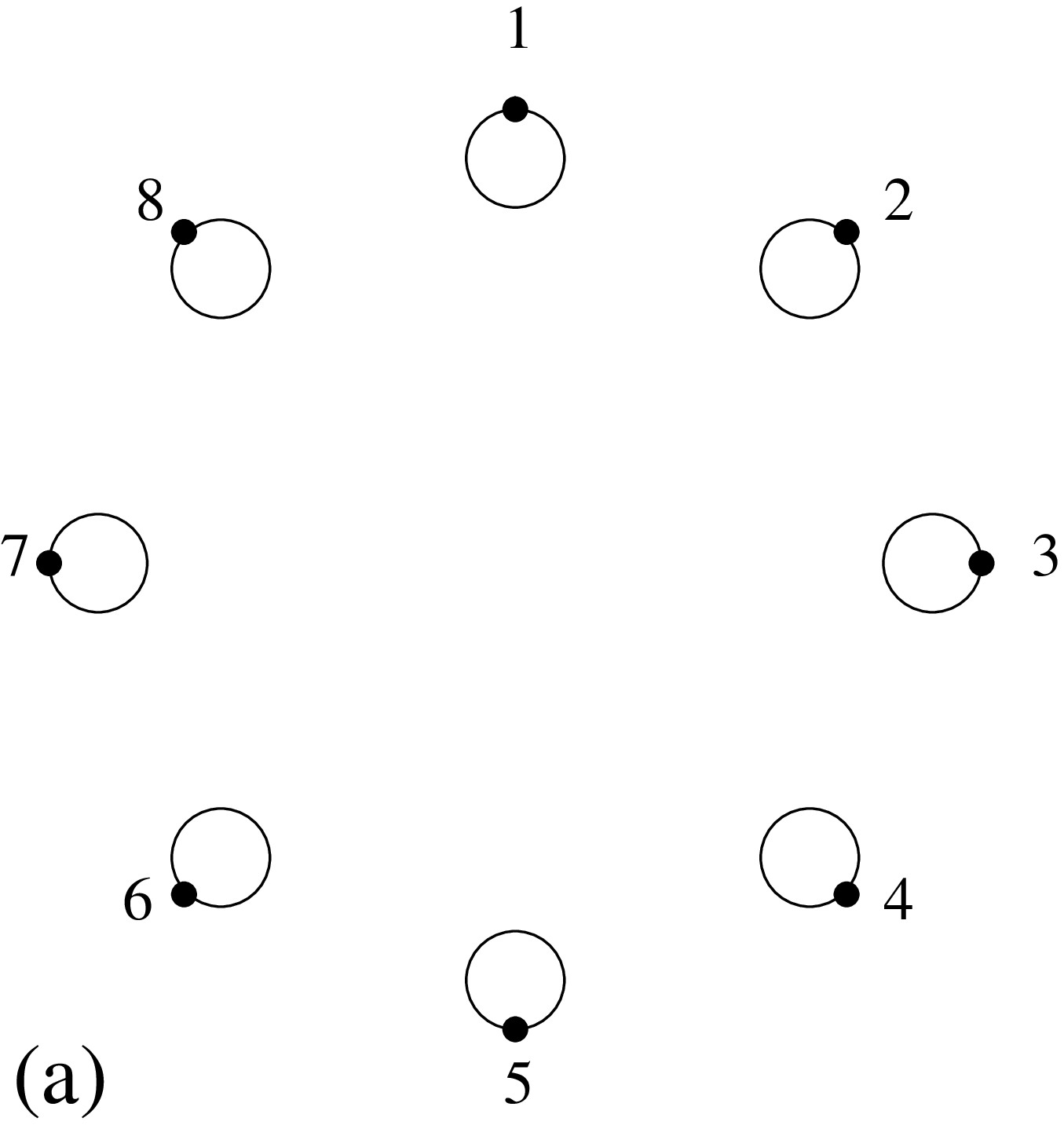}\hspace{3cm}
\includegraphics[scale=.4]{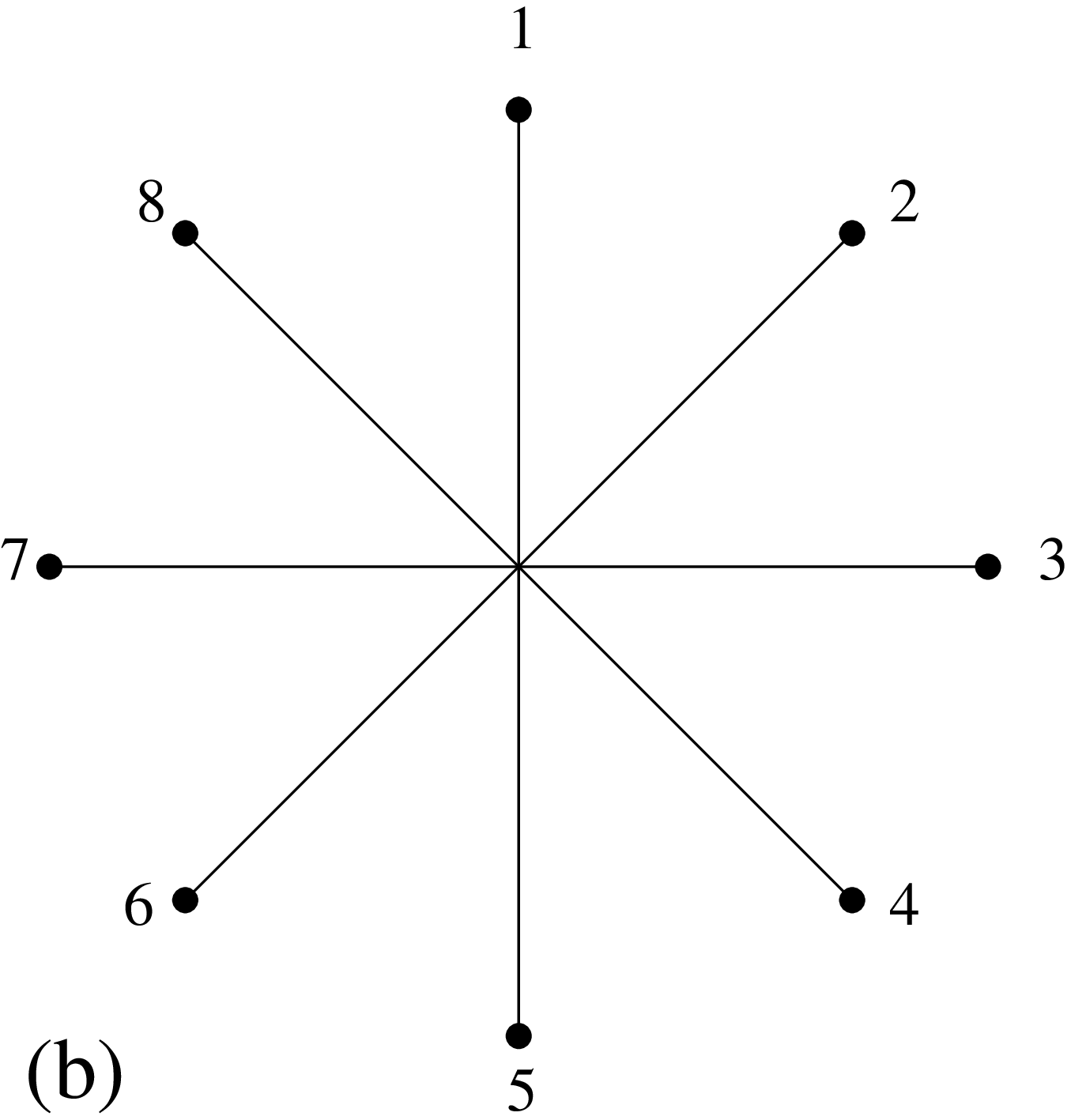}
\end{center}
\caption{Left panel (a): Representation of the group ``A''.
Right panel (b): Representation of the group ``A1''.  }\label{rep-A}
\end{figure}

(A) Absolute squares determined by $T^0_{I0}$, $T^c_{I0}$, $P^{z,0}_{I0}$, and
$P^{z,c}_{I0}$ for $I=0,1$, i.e., differential cross section and $z$-component of recoil
polarization for unpolarized and circularly polarized photons and unpolarized and polarized target: 
\beqa
  {\cal T}_{11}&=&
       \frac{1}{2\,c_{kin}}\Big(
              T^0_{ 00}+T^c_{ 00}
             +T^0_{ 10}+T^c_{ 10}
             +P^{ z,0}_{00}+P^{ z,c}_{00}
             +P^{ z,0}_{10}+P^{ z,c}_{10}
          \Big)\,,\\
  {\cal T}_{22}&=&
       \frac{1}{2\,c_{kin}}\Big(
              T^0_{ 00}+T^c_{ 00}
             +T^0_{ 10}+T^c_{ 10}
             -P^{ z,0}_{00}-P^{ z,c}_{00}
             -P^{ z,0}_{10}-P^{ z,c}_{10}
          \Big)\,,\\
  {\cal T}_{33}&=&
       \frac{1}{2\,c_{kin}}\Big(
              T^0_{ 00}-T^c_{ 00}
             +T^0_{ 10}-T^c_{ 10}
             +P^{ z,0}_{00}-P^{ z,c}_{00}
             +P^{ z,0}_{10}-P^{ z,c}_{10}
          \Big)\,,\\
  {\cal T}_{44}&=&
       \frac{1}{2\,c_{kin}}\Big(
              T^0_{ 00}-T^c_{ 00}
             +T^0_{ 10}-T^c_{ 10}
             -P^{ z,0}_{00}+P^{ z,c}_{00}
             -P^{ z,0}_{10}+P^{ z,c}_{10}
          \Big)\,,\\
  {\cal T}_{55}&=&
       \frac{1}{2\,c_{kin}}\Big(
              T^0_{ 00}+T^c_{ 00}
             -T^0_{ 10}-T^c_{ 10}
             +P^{ z,0}_{00}+P^{ z,c}_{00}
             -P^{ z,0}_{10}-P^{ z,c}_{10}
          \Big)\,,\\
  {\cal T}_{66}&=&
       \frac{1}{2\,c_{kin}}\Big(
              T^0_{ 00}+T^c_{ 00}
             -T^0_{ 10}-T^c_{ 10}
             -P^{ z,0}_{00}-P^{ z,c}_{00}
             +P^{ z,0}_{10}+P^{ z,c}_{10}
          \Big)\,,\\
  {\cal T}_{77}&=&
       \frac{1}{2\,c_{kin}}\Big(
              T^0_{ 00}-T^c_{ 00}
             -T^0_{ 10}+T^c_{ 10}
             +P^{ z,0}_{00}-P^{ z,c}_{00}
             -P^{ z,0}_{10}+P^{ z,c}_{10}
          \Big)\,,\\
  {\cal T}_{88}&=&
       \frac{1}{2\,c_{kin}}\Big(
              T^0_{ 00}-T^c_{ 00}
             -T^0_{ 10}+T^c_{ 10}
             -P^{ z,0}_{00}+P^{ z,c}_{00}
             +P^{ z,0}_{10}-P^{ z,c}_{10}
          \Big)\,.
\eeqa
The graphical representation is shown in the left panel (a)  of Fig.~\ref{rep-A}.

(A1) Interference terms determined by $U^0_{11}$, $U^c_{11}$,
$R^{z,0}_{11}$, and $R^{z,c}_{11}$, i.e., differential
cross section and $z$-component of recoil polarization for unpolarized
and circularly polarized photons and polarized target:
\beqa
  {\cal T}_{51}&=&-
       \frac{1}{2\sqrt{2}\,c_{kin}}\Big(
              U^0_{11}+U^c_{11}
             +R^{ z,0}_{11}+R^{ z,c}_{11}
          \Big)\,,\\
  {\cal T}_{62}&=&-
       \frac{1}{2\sqrt{2}\,c_{kin}}\Big(
              U^0_{11}+U^c_{11}
             -R^{ z,0}_{11}-R^{ z,c}_{11}
          \Big)\,,\\
  {\cal T}_{73}&=&-
       \frac{1}{2\sqrt{2}\,c_{kin}}\Big(
              U^0_{11}-U^c_{11}
             +R^{ z,0}_{11}-R^{ z,c}_{11}
          \Big)\,,\\
  {\cal T}_{84}&=&-
       \frac{1}{2\sqrt{2}\,c_{kin}}\Big(
              U^0_{11}-U^c_{11}
             -R^{ z,0}_{11}+R^{ z,c}_{11}
          \Big)\,.
\eeqa
The graphical representation is shown in the right panel (b)  of Fig.~\ref{rep-A}.

\begin{figure}[h]
\begin{center}
\includegraphics[scale=.4]{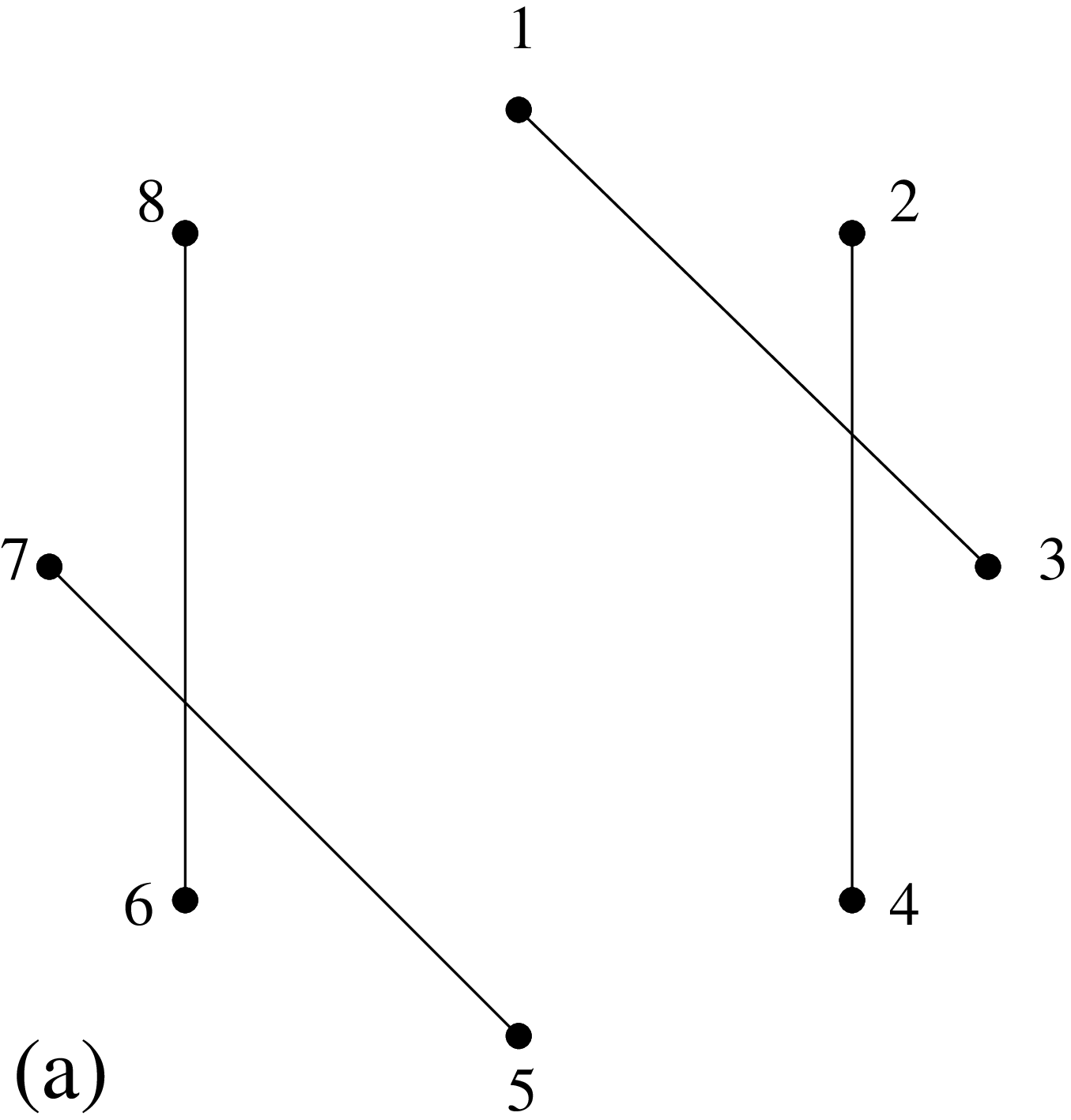}\hspace{3cm}
\includegraphics[scale=.4]{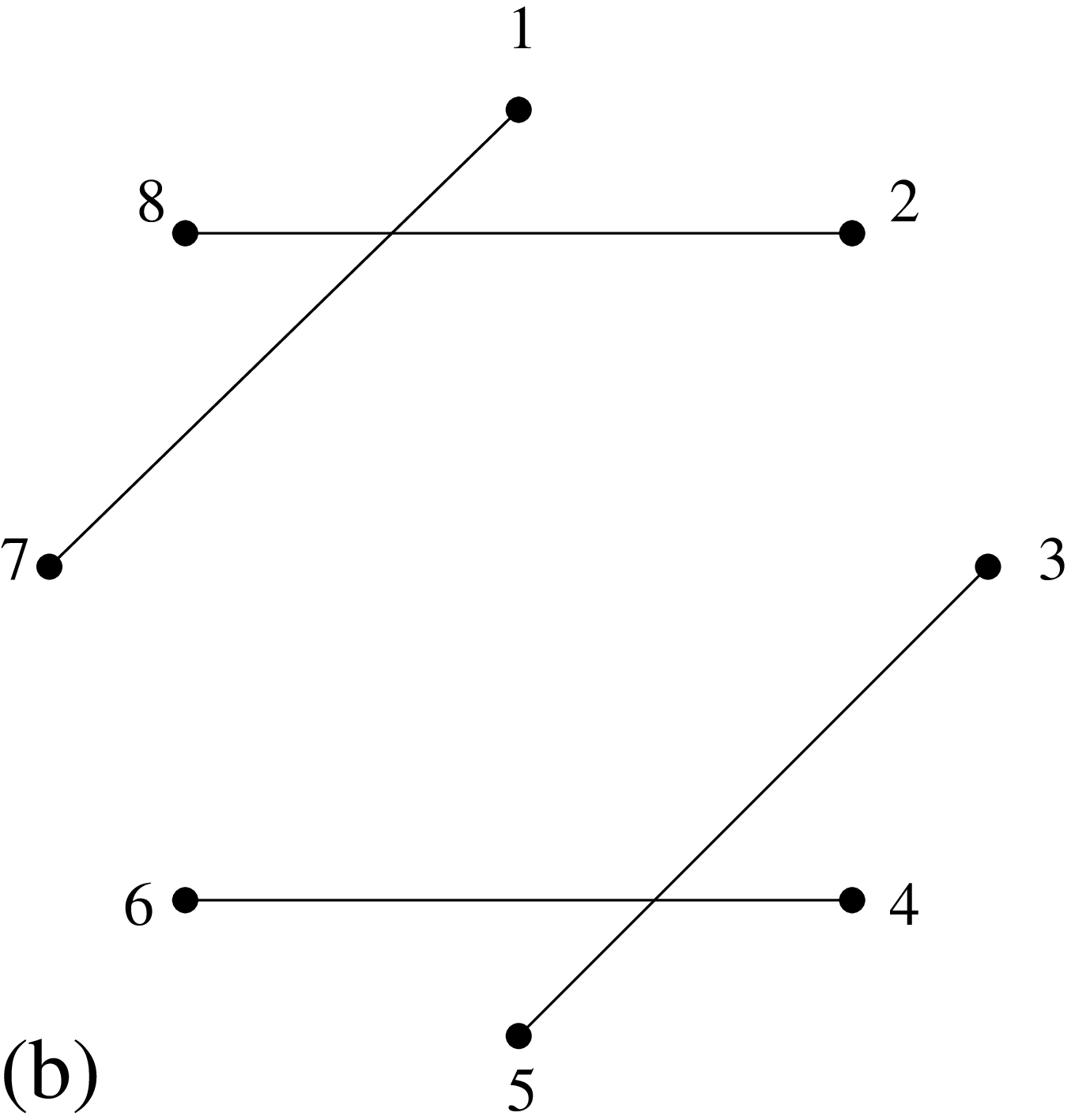}
\end{center}
\caption{Left panel (a): Representation of the group ``B''.
Right panel (b):  Representation of the group ``B1''. } \label{figB}
\end{figure}

(B) Interference terms determined by $U^\ell_{I0}$ and
$R^{z,\ell}_{I0}$ for $I=0,1$, i.e., differential
cross section and $z$-component of recoil polarization for linearly
polarized photons and unpolarized and polarized target:
\beqa
  {\cal T}_{13}&=&
      -\frac{1}{2\,c_{kin}}\Big(
              U^{\ell}_{ 0 0}
             +U^{\ell}_{ 1 0}
             +R^{ z,\ell}_{ 0 0}
             +R^{ z,\ell}_{ 1 0}
          \Big)\,,\\
  {\cal T}_{24}&=&
      -\frac{1}{2\,c_{kin}}\Big(
              U^{\ell}_{ 0 0}
             +U^{\ell}_{ 1 0}
             -R^{ z,\ell}_{ 0 0}
             -R^{ z,\ell}_{ 1 0}
          \Big)\,,\\
  {\cal T}_{57}&=&
      -\frac{1}{2\,c_{kin}}\Big(
              U^{\ell}_{ 0 0}
             -U^{\ell}_{ 1 0}
             +R^{ z,\ell}_{ 0 0}
             -R^{ z,\ell}_{ 1 0}
          \Big)\,,\\
  {\cal T}_{68}&=&
      -\frac{1}{2\,c_{kin}}\Big(
              U^{\ell}_{ 0 0}
             -U^{\ell}_{ 1 0}
             -R^{ z,\ell}_{ 0 0}
             +R^{ z,\ell}_{ 1 0}
          \Big)\,.
\eeqa
The graphical representation is shown in the left panel (a)  of Fig.~\ref{figB}.

(B1) Interference terms determined by $U^{\ell}_{1\pm 1}$ and
$R^{z,\ell}_{1\pm 1}$, i.e., differential
cross section and $z$-component of
recoil polarization for linearly
polarized photons and polarized target:
\beqa
  {\cal T}_{17}&=&-
       \frac{1}{\sqrt{2}\,c_{kin}}\Big(
              U^{\ell}_{ 1-1}
             +R^{ z,\ell}_{ 1-1}
          \Big)\,,\\
  {\cal T}_{28}&=&-
       \frac{1}{\sqrt{2}\,c_{kin}}\Big(
              U^{\ell}_{ 1-1}
             -R^{ z,\ell}_{ 1-1}
          \Big)\,,\\
  {\cal T}_{53}&=&
      \frac{1}{\sqrt{2}\,c_{kin}}\Big(
              U^{\ell}_{ 1 1}
             +R^{ z,\ell}_{ 1 1}
          \Big)\,,\\
  {\cal T}_{64}&=&
      \frac{1}{\sqrt{2}\,c_{kin}}\Big(
              U^{\ell}_{ 1 1}
             -R^{ z,\ell}_{ 1 1}
          \Big)\,.
\eeqa
The graphical representation is shown in the right panel (b)  of Fig.~\ref{figB}.

\begin{figure}[h]
\begin{center}
\includegraphics[scale=.4]{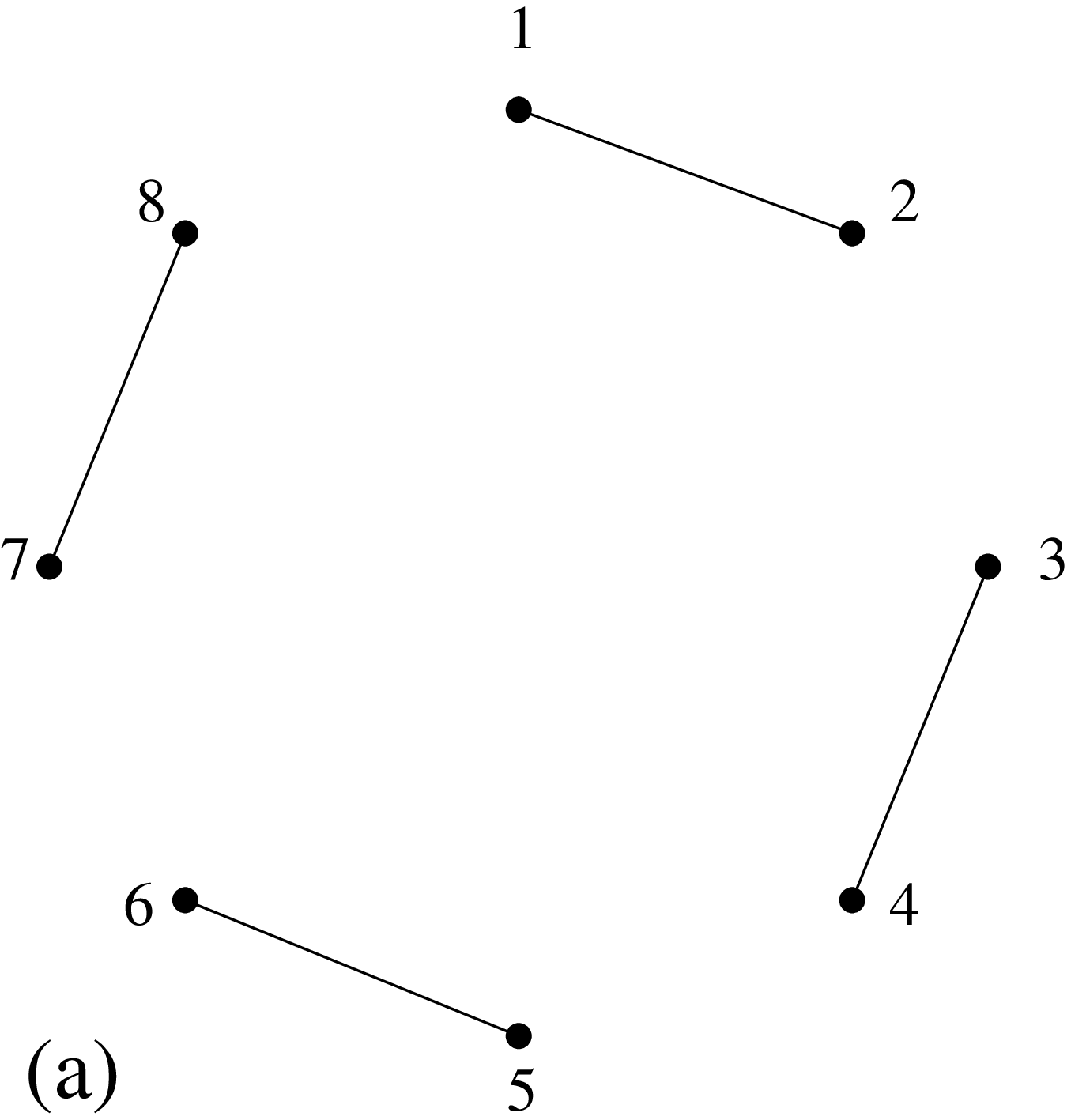}\hspace{3cm}
\includegraphics[scale=.4]{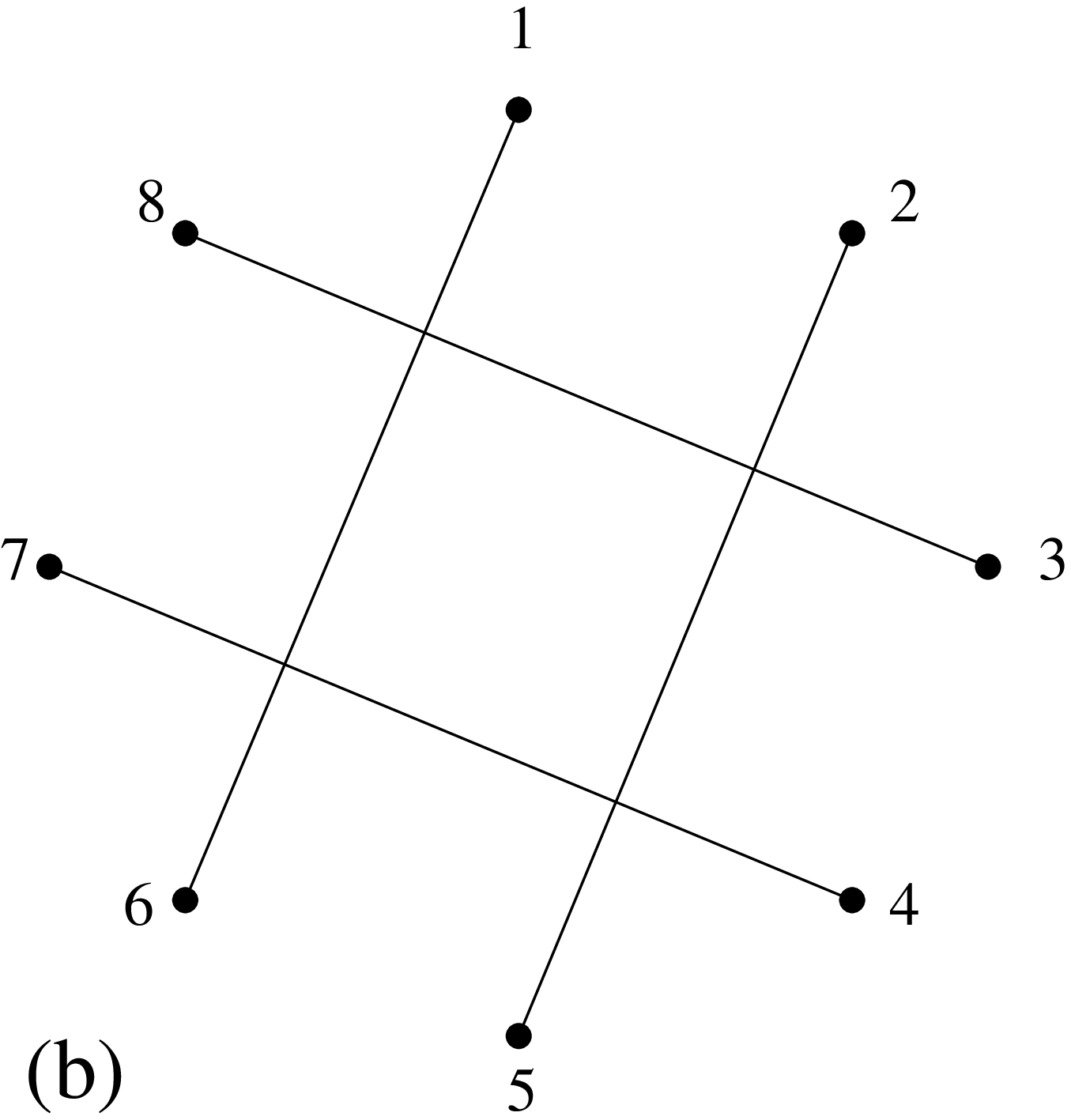}
\end{center}
\caption{Left panel (a): Representation of the group ``C''.
Right panel (b): Representation of the group ``C1''. } \label{figC}
\end{figure}

 (C) Interference terms determined by
$P^{1,0/c}_{I0}=-(P^{x,0/c}_{I0}+i\,P^{y,0}_{I0/c})/\sqrt{2}$ for
$I=0,1$, i.e., the transverse
spherical components of recoil polarization for unpolarized
and circularly polarized photons and unpolarized and polarized target:

\beqa
  {\cal T}_{12}&=&
      -\frac{1}{\sqrt{2}\,c_{kin}}\Big(
              P^{ 1,0}_{00}+P^{ 1,c}_{00}
             +P^{ 1,0}_{10}+P^{ 1,c}_{10}
          \Big)\,,\\
  {\cal T}_{34}&=&
      -\frac{1}{\sqrt{2}\,c_{kin}}\Big(
              P^{ 1,0}_{00}-P^{ 1,c}_{00}
             +P^{ 1,0}_{10}-P^{ 1,c}_{10}
          \Big)\,,\\
  {\cal T}_{56}&=&
      -\frac{1}{\sqrt{2}\,c_{kin}}\Big(
              P^{ 1,0}_{00}+P^{ 1,c}_{00}
             -P^{ 1,0}_{10}-P^{ 1,c}_{10}
          \Big)\,,\\
  {\cal T}_{78}&=&
      -\frac{1}{\sqrt{2}\,c_{kin}}\Big(
              P^{ 1,0}_{00}-P^{ 1,c}_{00}
             -P^{ 1,0}_{10}+P^{ 1,c}_{10}
          \Big)\,.
\eeqa
The graphical representation is shown in the left panel (a)  of Fig.~\ref{figC}.

(C1) Interference terms determined by
$R^{\pm 1,0/c}_{11}=\mp (R^{x,0/c}_{11}\pm i\,P^{y,0/c}_{11})/\sqrt{2}$,
i.e.\ the transverse spherical
component of recoil polarization for unpolarized
and circularly polarized photons and polarized target:

\beqa
  {\cal T}_{61}&=&-
       \frac{1}{2\, c_{kin}}\Big(
              R^{-1,0}_{11}+R^{-1,c}_{11}
          \Big)\,,\\
  {\cal T}_{52}&=&
      \frac{1}{2\, c_{kin}}\Big(
              R^{ 1,0}_{11}+R^{ 1,c}_{11}
          \Big)\,,\\
  {\cal T}_{83}&=&-
       \frac{1}{2\, c_{kin}}\Big(
              R^{-1,0}_{11}-R^{-1,c}_{11}
          \Big)\,,\\
  {\cal T}_{74}&=&
      \frac{1}{2\, c_{kin}}\Big(
              R^{ 1,0}_{11}-R^{ 1,c}_{11}
          \Big)\,.
\eeqa
The graphical representation is shown in the right panel (b)  of Fig.~\ref{figC}.

\begin{figure}[h]
\begin{center}
\includegraphics[scale=.4]{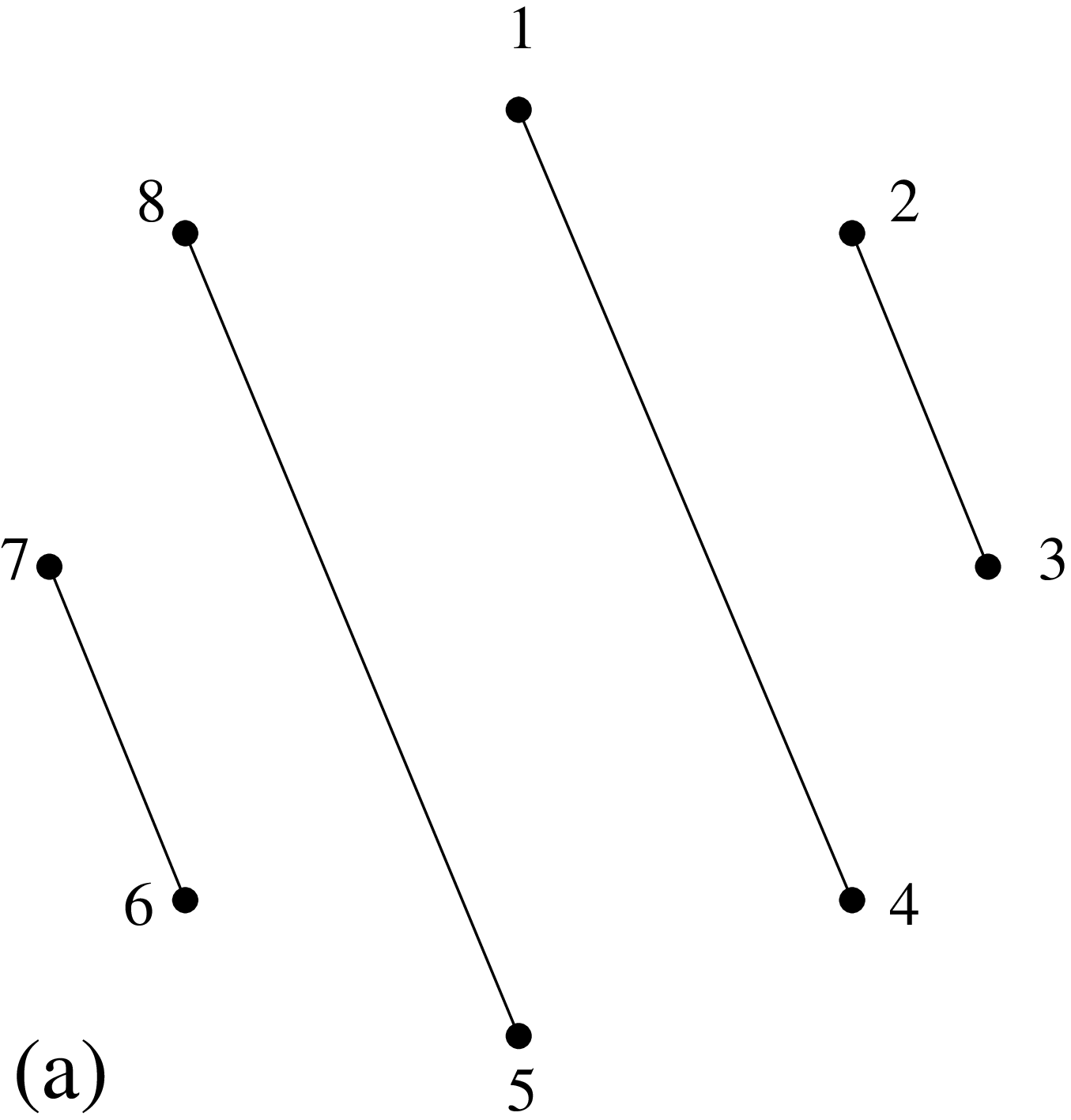}\hspace{3cm}
\includegraphics[scale=.4]{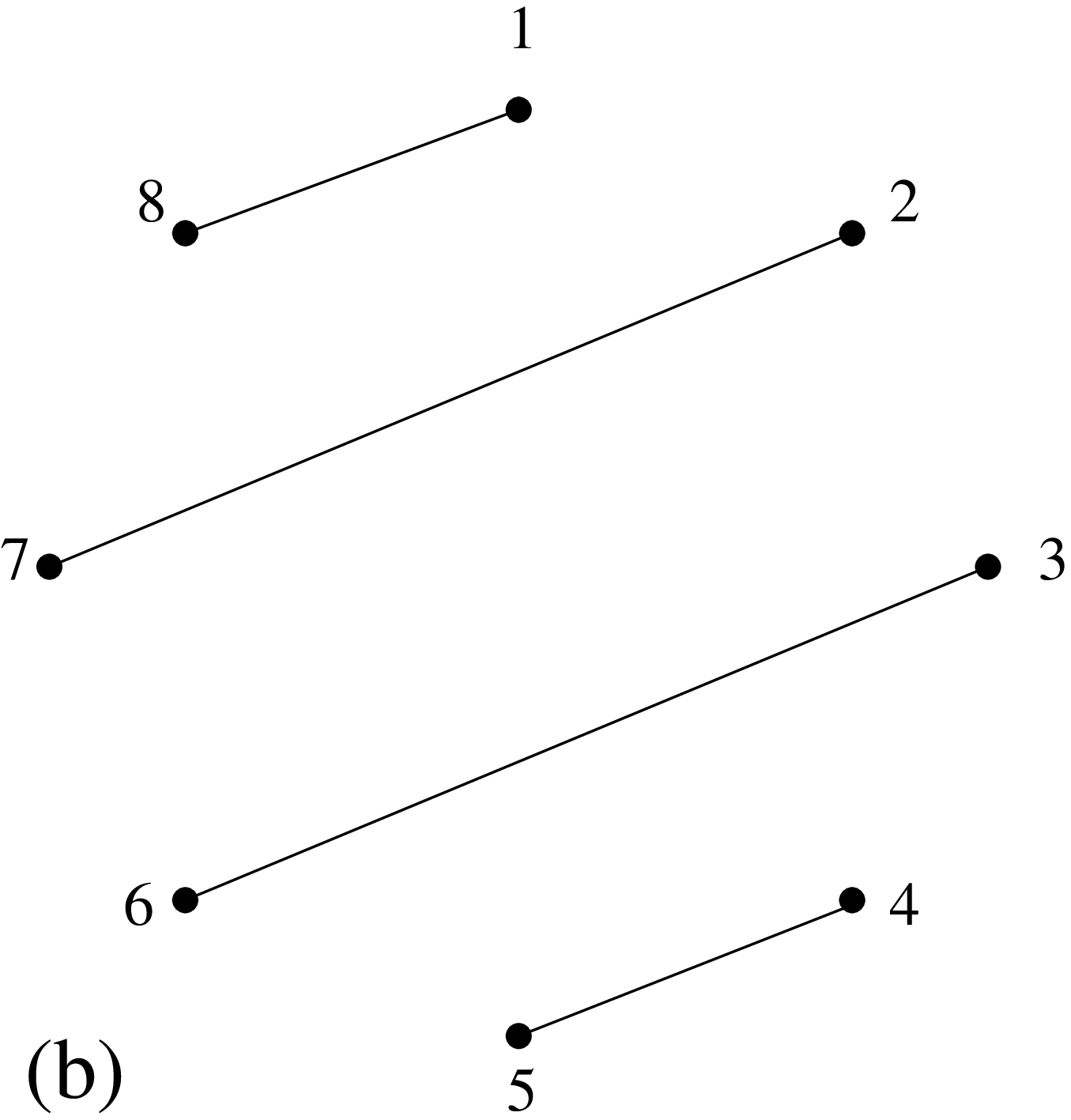}
\end{center}
\caption{Left panel (a): Representation of the group ``D''.
Right panel (b): Representation of the group ``D1''. } \label{figD}
\end{figure}

 (D) Interference terms determined by $R^{\pm 1,\ell}_{I0}$ for
 $I=0,1$, i.e., the transverse spherical components of recoil
polarization for linearly polarized photons and unpolarized and
polarized target:

\beqa
  {\cal T}_{14}&=&
       \frac{1}{\sqrt{2}\,c_{kin}}\Big(
              R^{ 1,\ell}_{ 0 0}
             +R^{ 1,\ell}_{ 1 0}
          \Big)\,,\\
  {\cal T}_{23}&=&
      -\frac{1}{\sqrt{2}\,c_{kin}}\Big(
              R^{-1,\ell}_{ 0 0}
             +R^{-1,\ell}_{ 1 0}
          \Big)\,,\\
  {\cal T}_{58}&=&
       \frac{1}{\sqrt{2}\,c_{kin}}\Big(
              R^{ 1,\ell}_{ 0 0}
             -R^{ 1,\ell}_{ 1 0}
          \Big)\,,\\
  {\cal T}_{67}&=&
      -\frac{1}{\sqrt{2}\,c_{kin}}\Big(
              R^{-1,\ell}_{ 0 0}
             -R^{-1,\ell}_{ 1 0}
          \Big)\,.
\eeqa
The graphical representation is shown in the left panel (a)  of Fig.~\ref{figD}.


(D1) Interference terms determined by $R^{\pm 1,\ell}_{1\pm 1}$, i.e.,
the transverse spherical components of recoil
polarization for linearly polarized photons and a
polarized target:

\beqa
  {\cal T}_{18}&=&
      \frac{1}{c_{kin}}
              R^{ 1,\ell}_{ 1-1}
          \,,\\
  {\cal T}_{27}&=&-
       \frac{1}{c_{kin}}
              R^{-1,\ell}_{ 1-1}
          \,,\\
  {\cal T}_{63}&=&
      \frac{1}{c_{kin}}
              R^{-1,\ell}_{ 1 1}
          \,,\\
  {\cal T}_{54}&=&-
       \frac{1}{c_{kin}}
              R^{ 1,\ell}_{ 1 1}
          \,.
\eeqa
The graphical representation is shown in the right panel (b)  of Fig.~\ref{figD}.

One should note that each group is represented by eight polarization observables, and
there is no overlap between the observables of the various groups, thus the total number
of 64 observables is evenly distributed over the eight groups. The four groups ``A'' through
``B1'' are associated with the observables of the differential cross section and recoil
polarization component $P_z$. The interference terms ${\cal T}_{j'j}$ of the groups
``A1'' through ``B1'' connect matrix elements with $(j',j)$ either both even or both odd.
The other four groups ``C'' through ``D1'' are associated with those of the recoil
polarization components $P_x$ and $P_y$. Here we have interference terms ${\cal T}_{j'j}$
with $j'$ even and $j$ odd or vice versa.

\renewcommand{\theequation}{C\arabic{equation}}
\setcounter{equation}{0}
\section{Construction of a complete set: First Method}
\label{methodA} In this appendix, we show how to express the matrix elements
$t_2,\dots,t_8$ by $t_1$ and the following observables \beqa T^0_{0 0}&=&
\frac{c_{kin}}{4}\,(
  {\cal T}_{11}  +{\cal T}_{22}  +{\cal T}_{33}  +{\cal T}_{44}
+{\cal T}_{55}  +{\cal T}_{66}  +{\cal T}_{77}  +{\cal T}_{88}
)\,,\label{t00}\\
U^0_{1 1}&=&T^0_{1 1}+i\,S^0_{1 1}=-
\frac{c_{kin}}{\sqrt{2}}\,(
  {\cal T}_{51}
  +{\cal T}_{62}
  +{\cal T}_{73}
  +{\cal T}_{84}
)\,,\label{u011}\\
U^c_{1 1}&=&T^c_{1 1}+i\,S^c_{1 1}=-
\frac{c_{kin}}{\sqrt{2}}\,(
  {\cal T}_{51}
  +{\cal T}_{62}
  -{\cal T}_{73}
  -{\cal T}_{84}
) \,,\\
R^{z,0}_{1 1}&=&P^{z,0}_{1 1}+i\,Q^{z,0}_{1 1}=-
\frac{c_{kin}}{\sqrt{2}}\,(
  {\cal T}_{51}
  -{\cal T}_{62}
  +{\cal T}_{73}
  -{\cal T}_{84}
) \,,\label{rz011}\\
R^{z,c}_{1 1}&=&P^{z,c}_{1 1}+i\,Q^{z,c}_{1 1}=-
\frac{c_{kin}}{\sqrt{2}}\,(
  {\cal T}_{51}
  -{\cal T}_{62}
  -{\cal T}_{73}
  +{\cal T}_{84}
)\,,\label{rzc11}
\\
U^\ell_{0 0}&=&T^\ell_{0 0}+i\,S^\ell_{0 0}=
-\frac{c_{kin}}{2}\,(
  {\cal T}_{13}
  +{\cal T}_{24}
  +{\cal T}_{57}
  +{\cal T}_{68}
)
\,,\label{ul00}\\
U^\ell_{1 0}&=&T^\ell_{1 0}+i\,S^\ell_{1 0}=
-\frac{c_{kin}}{2}\,(
  {\cal T}_{13}
  +{\cal T}_{24}
  -{\cal T}_{57}
  -{\cal T}_{68}
) \label{ul10}
\,,\\
P^{x+iy,0}_{0 0}&=&P^{x,0}_{0 0}+i\,P^{y,0}_{0 0}=
\frac{c_{kin}}{2}\,\Big(
  {\cal T}_{21}
  +{\cal T}_{43}
  +{\cal T}_{65}
  +{\cal T}_{87}
\Big)\,.\label{pxy}
\eeqa

From eqs.~(\ref{u011}) through (\ref{rzc11}) one first obtains
\beqa
{\cal T}_{51}&=&c_{51}=-\frac{1}{2\,\sqrt{2}\,
c_{kin}}(U^0_{1 1}+U^c_{1 1}
+R^{z,0}_{1 1}+R^{z,c}_{1 1})\,,\label{c51}\\
{\cal T}_{62}&=&c_{62}=-\frac{1}{2\,\sqrt{2}\,
c_{kin}}(U^0_{1 1}+U^c_{1 1}
-R^{z,0}_{1 1}-R^{z,c}_{1 1})\,,\label{c62}\\
{\cal T}_{73}&=&c_{73}=-\frac{1}{2\,\sqrt{2}\,
c_{kin}}(U^0_{1 1}-U^c_{1 1}
+R^{z,0}_{1 1}-R^{z,c}_{1 1})\,,\label{c73}\\
{\cal T}_{84}&=&c_{84}=-\frac{1}{2\,\sqrt{2}\,
c_{kin}}(U^0_{1 1}-U^c_{1 1}
-R^{z,0}_{1 1}+R^{z,c}_{1 1})\,. \label{c84}
\eeqa
This allows one to determine the matrix elements $t_j$ for $j=5,\dots
8$ from the ones for $j=1,\dots 4$.

As next, we will relate $t_j$ for
$j=2,\dots 4$ to $t_1$. To this end we will consider eqs.~(\ref{ul00})
and (\ref{ul10}) and obtain
\beqa
c^\ell_{+}&=&-\frac{1}{c_{kin}}\,(U^\ell_{00}+U^\ell_{10})={\cal T}_{13}
  +{\cal T}_{24}\,,\label{cl0}\\
c^\ell_{-}&=&-\frac{1}{c_{kin}}\,(U^\ell_{00}-U^\ell_{10})={\cal T}_{57}
  +{\cal T}_{68} \label{cl1}\,.
\eeqa
First, we express ${\cal T}_{24}$ by ${\cal T}_{13}$ using the obvious general
relation 
\beqa 
{\cal T}_{ab}=\frac{{\cal T}_{ac}{\cal T}_{db}}{{\cal T}_{dc}} \,,
\eeqa 
and insert into eq.~(\ref{cl1}) for ${\cal T}_{57}$ and ${\cal
  T}_{68} $ the following relations
\beq
{\cal T}_{57}=\frac{{\cal T}_{51}{\cal T}_{73}^*}{{\cal T}_{31}}
=\frac{c_{51}\,c_{73}^*}{{\cal T}_{31}}\,\quad\mathrm{and}\quad
{\cal T}_{68}=\frac{{\cal T}_{62}{\cal T}_{84}^*}{{\cal T}_{42}}
=\frac{c_{62}\,c_{84}^*}{{\cal T}_{42}}\,,
\eeq
yielding
\beq
\frac{c_{62}\,c_{84}^*}{{\cal T}_{42}}=c_{-}^\ell
-\frac{c_{51}\,c_{73}^*}{{\cal T}_{31}}\,.
\eeq
This allows one to express ${\cal T}_{24}$ by ${\cal T}_{13}$
\beq
{\cal T}_{24}=\frac{c_{62}^*\,c_{84}\,{\cal
    T}_{13}}{c^{\ell *}_{-}\,{\cal T}_{13}-c_{51}^*\,c_{73}}\,.
\eeq
With the help of this last relation one can eliminate ${\cal T}_{24}$
from eq.~(\ref{cl0}) resulting in a quadratic equation for ${\cal
  T}_{13}$
\beq
c^{\ell *}_{-}\,{\cal
  T}_{13}^2+(c_{62}^*\,c_{84}-c_{51}^*\,c_{73} -c^{\ell}_{+}\,c^{\ell
  *}_{-}) \,{\cal T}_{13}=-c^{\ell}_{+}c_{51}^*\,c_{73}\,,
\eeq
whose solution yields $t_3$ inverse proportional to $t_1^*$, i.e.\
\beq
c_{13}={\cal T}_{13}=\frac{1}{2}\,\Big(-a_3\pm\sqrt{a_3^2+4a_3b_3}\Big)\,,
\eeq
with
\beqa
a_3&=&\frac{1}{c^{\ell *}_{-}}\,\Big(c_{62}^*\,c_{84} -c_{51}^*\,c_{73}
-c^{\ell }_{+}\,c^{\ell *}_{-}\Big)\,,\\
b_3&=&-\frac{c^{\ell }_{+}\,c_{51}^*\,c_{73}}{c^{\ell *}_{-}}\,.
\eeqa
This is the first quadratic ambiguity one encounters. With $c_{13}={\cal
  T}_{13}$ known also ${\cal T}_{24}$ is found in terms of the
considered observables according to eq.~(\ref{cl0}), i.e.
\beq
c_{24}={\cal T}_{24}=c^{\ell }_{+}-c_{13}\,.
\eeq
Finally using first
\beq
{\cal T}_{34}=\frac{{\cal T}_{31}{\cal T}_{24}}{{\cal T}_{21}}
=\frac{c_{13}^*\,c_{24}}{{\cal T}_{21}}
\,,\quad
{\cal T}_{56}=\frac{{\cal T}_{51}{\cal T}_{62}^*}{{\cal T}_{21}}
=\frac{c_{62}^*\,c_{51}}{{\cal T}_{21}}\,,
\quad\mathrm{and}\quad
{\cal T}_{78}=\frac{{\cal T}_{84}^*{\cal T}_{73}}{{\cal T}_{24}^*{\cal
    T}_{13}}\,{\cal T}_{12}
=\frac{c_{84}^*\,c_{73}}{c_{24}^*\,c_{13}}\,{\cal T}_{12}\,,
\eeq
one obtains from eq.~(\ref{pxy})
\beq
c^{x+iy}=\frac{2}{c_{kin}}\,P^{x+iy,0}_{0 0}= {\cal T}_{12}^*
(1+\frac{{c}_{84}{c}_{73}^*}{c_{24}\,c_{13}^*})
+\frac{1}{{\cal T}_{12}}\,(c_{24}^*\,c_{13}
+{c}_{51}^*{c}_{62})\,,
\eeq
which is a quadratic equation for ${\cal T}_{12}$ of the type
\beq
|{\cal T}_{12}|^2+a_2\,{\cal T}_{12}=b_2
\eeq
with
\beq
a_2=-\frac{c^{x+iy}}{c_2}\,,\quad b_2=
-\frac{1}{c_2}(c_{24}^*\,c_{13}+c_{51}^*\,c_{62})
\,,\quad c_2=1+\frac{c_{73}^*c_{84}}{c_{24}\,c_{13}^*}\,.
\eeq
The solution reads with $c_{12}={\cal T}_{12}$
\beqa
\mathrm{Re}\,c_{12}&=&-\frac{1}{2}\,
\Big(\gamma\pm\sqrt{\gamma^2+4\gamma\delta}\Big)\,,\\
\mathrm{Im}\,c_{12}&=&\frac{1}{\mathrm{Re}\,a_2}\,
(\mathrm{Im}\,b_2
-\mathrm{Im}\,a_2\,\mathrm{Re}\, c_{12})\,,
\eeqa
where
\beqa
\gamma&=&\mathrm{Re} \,a_2
-2\frac{\mathrm{Im}\,a_2\mathrm{Im}\,b_2}{|a_2|^2}\,,\\
\delta&=&\frac{1}{|a_2|^2}\,\Big((\mathrm{Re}\,a_2)^2 \mathrm{Re}\,b_2
+(\mathrm{Re}\,a_2 \mathrm{Im}\,a_2-\mathrm{Im}\,b_2)
\mathrm{Im}\,b_2\Big)\,.
\eeqa
The quadratic solution for $\mathrm{Re}\,c_{12}$ introduces a second
ambiguity.

Thus all matrix elements $t_j$ for $j=2,\dots,8$ can be expressed by
$t_1$. In detail one finds $t_j=\frac{\sigma_j}{t_1^*}$ for $j=2,3,5,8$
with
\beq
\sigma_2=c_{12}\,,\quad \sigma_3=c_{13}\,,\quad
\sigma_5=c_{51}^* \,,\quad \sigma_8=\frac{c_{12}\,c_{84}^*}{c_{24}^*}\,,
\eeq
and $t_j=\tau_j\,t_1$ for $ j=4,6,7$ with
\beq
\tau_4=\frac {c_{24}}{c_{12}^*}\,,\quad
\tau_6=\frac {c_{62}^*}{c_{12}^*}\,,\quad
\tau_7=\frac {c_{73}^*}{c_{13}^*}\,.
\eeq

\renewcommand{\theequation}{D\arabic{equation}}
\setcounter{equation}{0}
\section{Construction of a complete set: Second Method}
\label{methodB}

In this appendix we will give some calculational details of the second
method.

i)  Elimination of the triple polarization observables
$R^{z,c}_{1 1}$ and $R^{z,\ell}_{1  0}$ using the
two quadratic relations in eqs.~(\ref{quadrela1}) and
(\ref{quadrelb1}):

Introducing for convenience the notation
\beqa
a_1=U^{\ell}_{0 0}+U^{\ell}_{1 0}+R^{z,\ell}_{0 0}\,,&&
b_1=(U^{0}_{1 1}+U^{c}_{1 1}+R^{z,0}_{1 1})\,,\\
a_2=U^{\ell}_{0 0}-U^{\ell}_{1 0}+R^{z,\ell}_{0 0}\,,&&
b_2=(U^{0}_{1 1}-U^{c}_{1 1}+R^{z,0}_{1 1})\,,\\
a_3=U^{\ell}_{0 0}+U^{\ell}_{1 0}-R^{z,\ell}_{0 0}\,,&&
b_3=(U^{0}_{1 1}+U^{c}_{1 1}-R^{z,0}_{1 1})\,,\\
a_4=U^{\ell}_{0 0}-U^{\ell}_{1 0}-R^{z,\ell}_{0 0}\,,&&
b_4=(U^{0}_{1 1}-U^{c}_{1 1}-R^{z,0}_{1 1})\,,
\eeqa
the two equations read
\beqa
2\,(a_1+R^{z,\ell}_{1 0})^*(a_2-R^{z,\ell}_{1 0})&=&(b_1+ R^{z,c}_{1 1}) (b_2- R^{z,c}_{1 1})^*\,,\label{quad1}\\
2\, (a_3-R^{z,\ell}_{1 0})^*(a_4+R^{z,\ell}_{1 0})&=&(b_3- R^{z,c}_{1 1}) (b_4+ R^{z,c}_{1 1})^*\,. \label{quad2} \eeqa 
Taking the sum and the difference one obtains 
\beqa
-2\, (a_1-a_3)^*R^{z,\ell}_{1 0}+2\,
(a_2-a_4)\,R^{z,\ell\,*}_{1 0} -4\,|R^{z,\ell}_{1 0}|^2
&=& -2\, a_1^*a_2-2\, a_3^*a_4 +b_1\,b_2^*+b_3\,b_4^* 
\nonumber\\
&&+(b_2-b_4)^*\, R^{z,c}_{1 1}-(b_1-b_3)\, R^{z,c\,*}_{1 1}
-2\,|R^{z,c}_{1 1}|^2\,, 
\label{sum}\\
-2\, (a_1+a_3)^*R^{z,\ell}_{1 0}+2\,
(a_2+a_4)\,R^{z,\ell \,*}_{1 0} 
&=&  -2\, a_1^*a_2+2\, a_3^*a_4 +b_1\,b_2^*-b_3\,b_4^* 
\nonumber\\
&&+(b_2^*+b_4^*)\, R^{z,c}_{1 1}-(b_1+b_3)\, R^{z,c\,*}_{1 1}
\,. \label{diff}
\eeqa 
The latter is a linear equation between $R^{z,\ell}_{1 0}$ and $
R^{z,c}_{1 1}$, which reads explicitely
\beq
-2\,(U_{00}^{\ell}+U_{10}^{\ell})^*\,R^{z,\ell}_{1 0}
+2\,(U_{00}^{\ell}-U_{10}^{\ell})\,R^{z,\ell\,*}_{1 0}=
 (U_{11}^{0}-U_{11}^{c})^*\, R^{z,c}_{1 1}
-(U_{11}^{0}+U_{11}^{c}) \, R^{z,c\,*}_{1 1}+\varepsilon\,
\eeq
with
\beqa
\varepsilon&=&-2\,[R^{z,\ell\,*}_{00}(U_{00}^{\ell}-U_{10}^{\ell})
+R^{z,\ell}_{00}(U_{00}^{\ell\,*}+U_{10}^{\ell\,*})]
+[R^{z,c}_{11}(U_{11}^{0\,*}-U_{11}^{c\,*})
+R^{z,c\,*}_{11}(U_{11}^{0}+U_{11}^{c})]
\,.
\eeqa
Thus one can eliminate $R^{z,\ell}_{1 0}$ by relating it to 
$R^{z,c}_{1 1}$ in the form 
$R^{z,\ell}_{1 0}=x\, R^{z,c}_{1 1}+y\, R^{z,c\,*}_{1 1}+z$. 
Explicitly one finds
\beqa
R^{z,\ell}_{1 0}&=& 
\frac{1}{4\,\mathrm{Re}(U_{00}^{\ell}U_{10}^{\ell *})}\,
\Big((U_{00}^{\ell}U_{11}^{c\,*} 
-U_{10}^{\ell}U_{11}^{0\,*}) \, R^{z,c}_{1 1}
+ (U_{00}^{\ell}U_{11}^{c} 
+U_{10}^{\ell}U_{11}^{0})\, R^{z,c\,*}_{1 1}\nonumber\\
&&+4\,R^{z,\ell\,*}_{00}U_{00}^{\ell}
-U_{00}^{\ell} \mathrm{Re}(R^{z,0\,*}_{11}U_{11}^{0})
-2\,i\,U_{10}^{\ell} \mathrm{Im}(2\,R^{z,\ell\,*}_{00}U_{10}^{\ell}
+R^{z,0\,*}_{11}U_{11}^{c} )\Big)
\,.
\eeqa

Finally, for the elimination of $R^{z,c}_{1 1}$ one can use eq.~(\ref{sum}).
Firstly, its imaginary part yields a linear equation between the real
and the imaginary part of $R^{z,c}_{1 1}$, i.e.
\beqa
2\,\mathrm{Im}(R^{z,\ell}_{00}\,R^{z,\ell\,*}_{10})
&=&
\mathrm{Im}(R^{z,0\,*}_{11}\,R^{z,c}_{11})
+\mathrm{Im}(U_{11}^{0\,*}\,U_{11}^{c}
-2\,U_{10}^{\ell\,*}\,U_{00}^{\ell})\,.
\eeqa
It allows the elimination of $\mathrm{Im}\,R^{z,c}_{1 1}$. 
For the elimination of the remaining
real part $\mathrm{Re}\,R^{z,c}_{1 1}$ one can utilize the real part
of eq.~(\ref{sum}) which takes the simple form of a quadratic equation
in $\mathrm{Re}\,R^{z,c}_{1 1}$:
\beq
2\,|R^{z,c}_{1 1}|^2-4\,|x\, R^{z,c}_{1 1}+y\, R^{z,c\,*}_{1 1}+z|^2=
|U_{11}^{0}|^2+|U_{11}^{c}|^2 +|R^{z,0}_{1 1}|^2
-(|U_{00}^{\ell}|^2+|U_{10}^{\ell}|^2+|R^{z,\ell}_{00}|^2)\,,
\eeq
resulting in another quadratic ambiguity.

ii)  Elimination of the polarization observables $P^{1,0}_{1 0}$, 
$P^{1,c}_{0 0}$, and $P^{1,c}_{1 0}$ using the three relations
in eqs.~(\ref{quadrelc}), (\ref{quadreld}), and (\ref{quadrelg}):

In order to simplify the notation, we introduce for convenience
\beq
a=P^{1,0}_{0 0}\,,\quad x=P^{1,c}_{0 0}\,,
\quad y=P^{1,0}_{1 0}\,,\quad z=P^{1,c}_{1 0}\,.
\eeq
The three equations then read in the following form
\beqa
(a+x+y+z)^*(a-x+y-z)&=&c_1\,,\label{eq1}\\
(a+x-y-z)^*(a-x-y+z)&=&c_2\,,\label{eq2}\\
(a+x+y+z)^*(a+x-y-z)&=&c_3\,, \label{eq3}
\eeqa
where
\beqa
c_1&=&\frac{1}{2}\,
(U^{\ell}_{0 0}+U^{\ell}_{1 0}+R^{z,\ell}_{0 0}+R^{z,\ell}_{1 0})^* 
(U^{\ell}_{0 0}+U^{\ell}_{1 0}-R^{z,\ell}_{0 0}-R^{z,\ell}_{1 0})\,,\\
c_2&=&\frac{1}{2}\,
(U^{\ell}_{0 0}-U^{\ell}_{1 0}+R^{z,\ell}_{0 0}-R^{z,\ell}_{1 0})^* 
(U^{\ell}_{0 0}-U^{\ell}_{1 0}-R^{z,\ell}_{0 0}+R^{z,\ell}_{1 0}) \,,\\
c_3&=&\frac{1}{2}\,
(U^0_{11}+U^c_{11}-R^{ z,0}_{11}-R^{ z,c}_{11})^*
(U^{\ell}_{ 0 0}-U^{\ell}_{ 1 0}+R^{ z,\ell}_{ 0 0}-R^{ z,\ell}_{ 1 0})
\frac{U^{\ell}_{ 0 0}+U^{\ell}_{ 1 0}+R^{ z,\ell}_{ 0 0}+R^{ z,\ell}_{ 1 0}}
{(U^0_{11}-U^c_{11}+R^{ z,0}_{11}-R^{ z,c}_{11})^*}\,.
\eeqa
Dividing eq.~(\ref{eq1}) and the complex conjugate
of eq.~(\ref{eq2}) by eq.~(\ref{eq3}) yields two linear equations, i.e.
\beqa
(a-x+y-z)&=&\frac{c_1}{c_3}\,(a+x-y-z)\,,\\
(a-x-y+z)^*&=&\frac{c_2^*}{c_3}\,(a+x+y+z)^*\,,
\eeqa
from which $x$ and $y$ can be related to $z$ according to
\beqa
x=\alpha_x\,z+\beta_x\,,\quad y=\alpha_y\,z+\beta_y\,,
\eeqa
with
\beq
\alpha_x=\frac{1}{2}\Big(\frac{c_1-c_3}{c_1+c_3}
+\frac{c_3^*-c_2}{c_3^*+c_2}\Big)\,,\quad
\alpha_y=\frac{1}{2}\Big(-\frac{c_1-c_3}{c_1+c_3}
+\frac{c_3^*-c_2}{c_3^*+c_2}\Big) \,,\quad \beta_x=\alpha_y\,a \,,\quad \beta_y=\alpha_x\,a\,.
\eeq
In order to determine $z$ we take the sum of eqs.~(\ref{eq1})
and (\ref{eq3}), resulting in
\beq
(a+x+y+z)\,(a-z)^*=\frac{1}{2}\,(c_1+c_3)^*\,.
\eeq
Insertion of the expressions for $x$ and $y$ yields a quadratic
equation for $z$
\beq
\frac{a+\beta_x+\beta_y}{1+\alpha_x+\alpha_y}\,(a-z)^*
+a^*z-|z|^2=\frac{(c_1+c_3)^*}{2\,(1+\alpha_x+\alpha_y)}\,,
\eeq
which is solved easily. 


\begin{thebibliography}{00}
\bibitem{RoO05}
W.~Roberts and T.~Oed, Phys.\ Rev.\ C {\bf 71}, 055201 (2005).

\bibitem{FiA13}
A.~Fix and H.~Arenh\"ovel, Phys.\ Rev.\ C {\bf 87}, 045503 (2013).

\bibitem{FiA12}
A.~Fix and H.~Arenh\"ovel, Phys.\ Rev.\ C {\bf 85}, 035502 (2012).

\bibitem{FiA11}
A.~Fix and H.~Arenh\"ovel, Phys.\ Rev.\ C {\bf 83}, 015503 (2011).

\bibitem{ArLT98}
H. Arenh\"ovel, W. Leidemann, and E.L. Tomusiak, Nucl. Phys. A {\bf
  641}, 517 (1998).

\bibitem{ArLT00}
H.~Arenh\"ovel, W.~Leidemann and E.~L.~Tomusiak, Few Body Syst.\  {\bf
  28}, 147 (2000)

\end{thebibliography}
\end{document}